\documentclass[trackchanges,twocolumn,tighten]{aastex7}
\usepackage{longtable}
\usepackage{placeins}
\usepackage{threeparttable}
\usepackage{scalerel}
\usepackage{amsmath}
\usepackage{float}
\usepackage{graphicx}
\usepackage{mathptmx}
\usepackage{txfonts}
\usepackage[T1]{fontenc}
\usepackage{microtype}
\usepackage{csquotes}
\usepackage{ae,aecompl}
\usepackage{etoolbox}
\apptocmd{\thebibliography}{\raggedright}{}{}

\begin{document}

\title{Is Milky Way gravitationally stable? A \texttt{TNG50} view from cosmic noon to the present day}

\author[orcid=0000-0002-1250-4359,sname='K. Aditya']{K. Aditya}
\affiliation{Raman Research Institute, C. V. Raman Avenue, 5th Cross Road, Sadashivanagar, Bengaluru, 560080, India}\email[show]{kaditya.astro@gmail.com}

\author[orcid=0000-0003-3657-0200 , sname='Sandeep Kataria']{Sandeep Kataria}
\affiliation{Space, Planetary \& Astronomical Sciences \& Engineering (SPASE), IIT Kanpur, 208016, India}
\email[show]{skkataria.iit@gmail.com }

\begin{abstract}
We investigate the stability of Milky Way analogs (MWAs) in the \texttt{TNG50} simulation against the growth of local 
axisymmetric instabilities, tracing their evolution from cosmic noon ($z=2.5$) to the present day ($z=0$). 
Using a two-component stability criterion that accounts for stars, gas, and the force field of the dark matter halo, 
we compute the net stability parameter ($Q_{T}$), the critical gas surface density ($\Sigma_{c}$), and the 
instability timescale ($\tau$) for 10 barred and 10 unbarred MWAs. We find that these galaxies remain 
stable to axisymmetric instabilities at all epochs, with $Q_{T}^{\min}>2$. The stability levels increase toward 
higher redshift, where enhanced gas velocity dispersion counterbalances the destabilizing effect of larger gas 
fractions. Further, the barred MWAs consistently show lower $Q_{T}^{\min}$ than unbarred ones. The gas density 
remains subcritical ($\Sigma_{g}<\Sigma_{c}$) across radii and epochs, implying that local axisymmetric instabilities 
are not the primary channel for star formation. Growth timescales are short (a few Myr) in central regions 
but increase exponentially to several Gyr in the outer disc, naturally explaining the concentration of star formation 
toward galactic centers. We study the effect of gas dissipation and turbulence in ISM and find that while MWAs are stable 
against axisymmetric instabilities $(Q_{T}>1)$, a combination of gas dissipation and turbulence in ISM can 
destabilize the disc at small scales even when $Q_{T}>1$.
\end{abstract}


\section{Introduction}
Local gravitational instabilities play a central role in various galaxy formation and evolution processes, and have been investigated since the pioneering works of \cite{safronov1960gravitational, toomre1964gravitational, goldreich1965ii}. The theoretical models of disc instabilities allow us to quantify the dynamical state of the system through a single number often called \emph{disc instability criterion}, first proposed for a self-gravitating disc of stars by \cite{toomre1964gravitational}. This criterion over the years has been developed and modified to describe self-consistently the dynamical and structural information available from observations, for eg: stability of disc of stars \citep{toomre1964gravitational}, stability of gas disc \cite{goldreich1965ii}, gas + stellar disc \citep{jog1984galactic, jog1996local,romeo2011effective}, multiple self-gravitating disc components \citep{rafikov2001local,romeo2013simple}, effect of dark matter \citep{jog2014effective, aditya2024does}, effect of disc thickness \citep{meidt2022phangs,nipoti2024local} and effect of turbulence \citep{hoffmann2012effect,shadmehri2012gravitational}. Thus, the current theoretical models of local gravitational instabilities encompass the complete mass inventory of the galaxy, including multiple stellar and gas components, their dark matter halo, and kinematic information through each component's velocity dispersions and rotation velocity.

The theory of local instabilities is closely connected to the observed star formation; since the instabilities drive the fragmentation of the gas disc into gas clumps, which eventually form stars \cite{kennicutt1989star,wang1994gravitational,bournaud2007rapid,agertz2009disc, dekel2009formation,krumholz2018unified}, with more stable discs exhibiting a lower star formation rate \cite{aditya2023stability}. The stability criterion also explains the morphological differences in the nearby galaxies, with irregular galaxies exhibiting higher stability levels than the spirals \citep{aditya2023stability}. Similarly, low surface brightness galaxies and superthin galaxies \citep{10.1093/mnras/stab155,aditya2022h} exhibit an overall higher net stability level than nearby disc galaxies. Recent high-resolution multi-wavelength observations are beginning to probe how instabilities regulate structure on smaller, sub-galactic scales. For example, studies based on galaxies from the PHANGS (Physics at High Angular Resolution in Nearby GalaxieS) survey \citep{schinnerer2019physics} show that the multi-scale filamentary structure observed in their gas discs aligns with the critical length scales predicted by turbulent Jeans instabilities \citep{meidt2023phangs}. 

At higher redshifts, observational studies have shown that galaxies are prone to local gravitational instabilities \citep{genzel2017strongly, walter2022alma, rizzo2020dynamically, aditya2023stability, bacchini20243d}. Recent James Webb Space Telescope observations have revealed that disc galaxies were already assembling in the early universe, within the first few billion years after the Big Bang \citep{smethurst2025galaxy, gillman2024structure}. Although the galaxies observed in the early universe were precursors to the galaxies in the local universe, it is not straightforward to make a one-to-one mapping of the nearby galaxies and their progenitors observed in the early universe.
Cosmological simulations, such as \texttt{TNG50} $(51.7 Mpc^{3})$ \citep{nelson2019illustristng}, provide us with an opportunity to make a one-to-one connection between galaxies observed in the local universe and their high-redshift progenitors, enabling us to study their continuous evolution. In the present study, we utilize a sample of Milky Way Analogs (MWAs) \citep{pillepich2024milky} from the \texttt{TNG50} suite of simulations to address the following questions regarding the dynamical state of the galaxy at various stages of its evolution, from cosmic noon to the present day.

\begin{enumerate}
    \item How do stability levels change across the galaxy, and are stability levels driven by stars or by gas?
    \item How does the critical gas surface density vary spatially, and how does it relate to the actual gas surface density?
    \item How does the timescale for the growth of instabilities vary across the galaxy?
    \item How is star formation distributed across the galaxy disc?
\end{enumerate}

We will present our sample of MW analogs from \texttt{TNG50} simulations in \S 2 and the dynamical model of disc instabilities in \S 3. We will present the results from our work in \S 4, discuss the implications of our results in \S 5, and summarize our results in \S 6.

\section{Milky Way analogs from \texttt{TNG50} simulations}
 We select our sample from the \texttt{TNG50} suite of simulations. The \texttt{TNG50} suite of simulations has the smallest volume among the \texttt{TNG} suite of simulations but offers the highest resolution. It evolves a (50 Mpc)$^3$ comoving volume, sampled with $2160^3$ dark matter particles and $2160^3$ initial gas cells \citep{nelson2019illustristng, pillepich2019first}, achieving a baryonic mass resolution of $\sim8.5 \times 10^4\ M_\odot$ and a dark matter particle mass resolution of $\sim4.5 \times 10^5\ M_\odot$. \texttt{TNG50} employs the moving-mesh code AREPO \citep{weinberger2020arepo} and incorporates the IllustrisTNG galaxy formation model, as described in detail by \citet{pillepich2018first} and \citet{weinberger2018supermassive}. The simulation solves the coupled equations of gravity and magnetohydrodynamics in an expanding Universe, while also modeling key baryonic processes, including radiative cooling and heating, star formation, stellar evolution, and chemical enrichment, as well as feedback from stars and supermassive black holes (SMBHs), including their seeding and growth. The gas in the TNG simulations is modeled with subgrid physics of multiphase ISM having cold (dense clouds where stars form) and hot (diffuse gas heated by supernova) gas components  \citep{Springel.Hernquist.2003}). The threshold criteria for star formation are reached when the density of gas $(n_H) \geq 0.13 \hspace{1mm} cm^{-3}$. The model is tuned to produce the Kennicutt-Schmidt relation \citep{kennicutt1998star} and controls the star formation rate accordingly. The initial conditions were set at redshift $z = 127$, assuming a cosmology consistent with the Planck 2015 results \citep{planck2016ade}. The stellar particles in \texttt{TNG50} simulations represent simple mono-age stellar populations, characterized by an initial mass function (IMF) following \citet{chabrier2003galactic}. \texttt{TNG50} is particularly well-suited for studying disc instabilities due to its high mass and spatial resolution \citep{nelson2019illustristng, pillepich2019first}.
 At $z = 0$, the gas cells in star-forming regions of MWAs range from 50–200 pc, with an average value of 150 pc. It is also important to note that in the inner regions
 ($R$ $<$ 1 kpc), the gas cell sizes are typically $<$ 100 pc, which is smaller than the gravitational softening length of the stellar and dark matter particles. This mismatch in resolution
  can lead to reduced coupling between the gas and stellar components. The gravitational potential is softened on resolution-element-dependent scales: stellar and dark matter particles have a softening length of 288 pc, and gas cells down to 72 pc. These resolutions are sufficient to robustly derive key quantities such as stellar and gas surface density profiles, velocity dispersion profiles, and the total gravitational potential. These are crucial ingredients for computing the local stability of disc galaxies against axisymmetric instabilities. We select our sample from the \texttt{TNG50} Milky Way analog catalog \citep{pillepich2024milky}. The catalog contains 198 Milky Way/M31 analogs, characterized by disky morphologies, stellar masses in the range $M_\ast = 10^{10.5} - 10^{11.2}\ M_\odot$, and residing in Milky Way–like environments at $z = 0$. The disc scalelengths of the barred galaxies in our sample span 3.2–7.1 kpc, while those of the unbarred galaxies range from 2.8 to 11.3 kpc.

\begin{figure*} 
\centering
\includegraphics[scale=0.6]{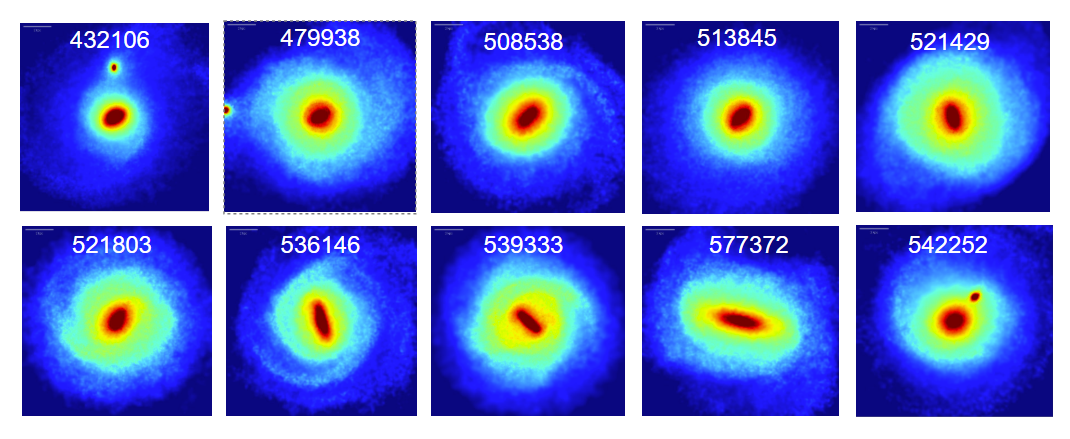}
\includegraphics[scale=0.6]{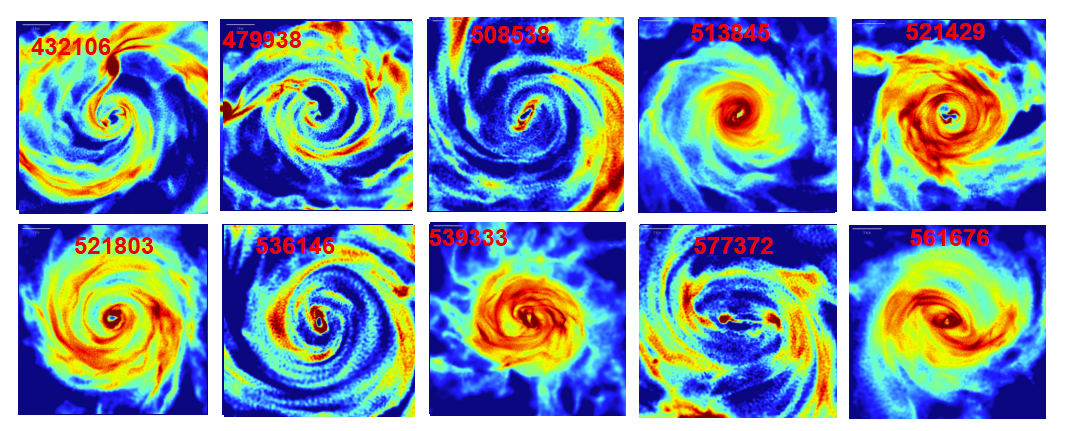}
\caption{Face-on maps of stellar density [top] and gas density [bottom] for barred galaxies at $z=0$. The red color corresponds to the highest density, and the blue color corresponds to the lowest density.}
\label{fig:barred_star}
\end{figure*}

\begin{figure*}
\centering
\includegraphics[scale=0.6]{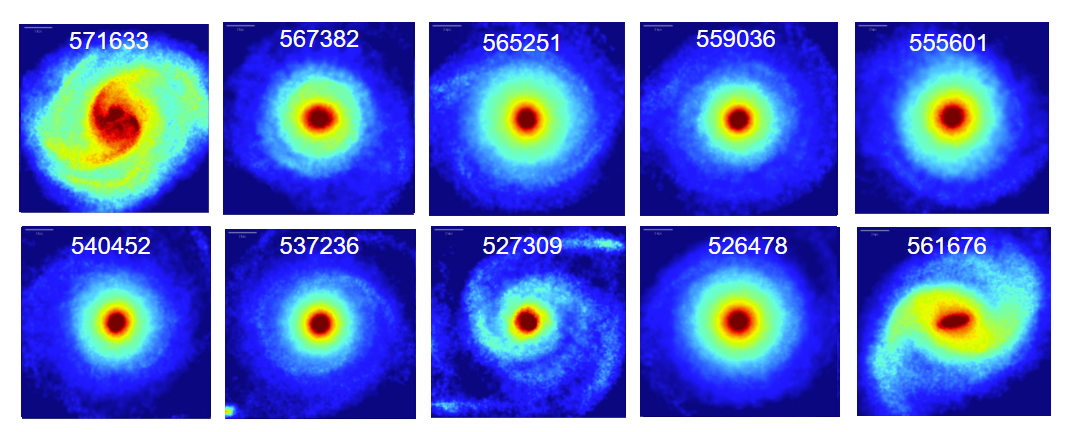}
\includegraphics[scale=0.6]{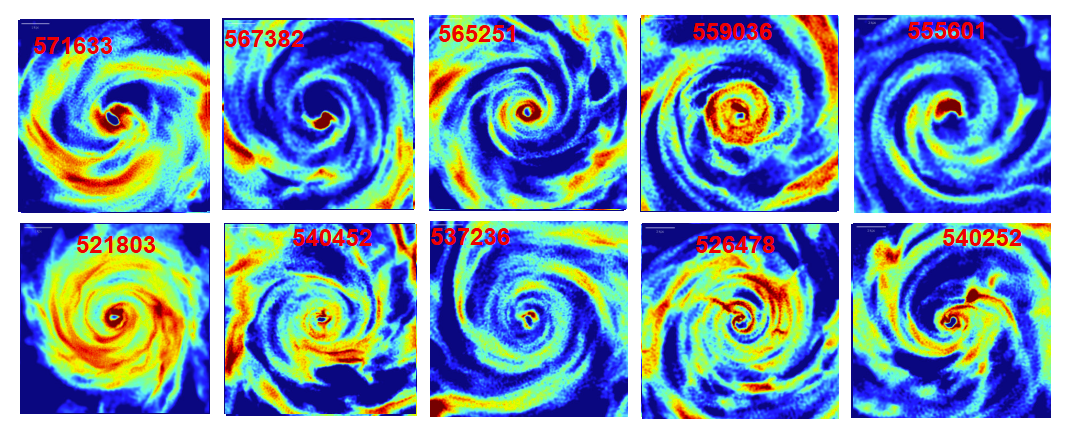}
\caption{Face-on maps of stellar density [top] and gas density [bottom] for unbarred galaxies at $z=0$. The red color corresponds to the highest density, and the blue corresponds to the lowest density.}
\label{fig:unbarred_star}
\end{figure*}

 We select a small representative subset of 10 barred and 10 unbarred MWAs from the parent catalog, all with similar dynamical masses, for our analysis. The division into barred and unbarred galaxies is based on classification provided by \cite{2020MNRAS.491.2547R,2022MNRAS.512.5339R, Zana_2022}. We select stars and gas by applying a simple geometric cutoff between $-1.5$ kpc $< z<$ 1.5 kpc and $R>$1 kpc at all epochs. The simple geometric cutoff allows us to probe and compare the same spatial region of the galaxy at all epochs. To assess the local stability of galaxies against the growth of gravitational instabilities, we need to measure the surface densities of stellar and gas discs, the radial velocity dispersion of stars and gas, and the circular velocity of stars, gas, and the dark matter halo. We use the publicly available package \texttt{pynbody} \citep{pontzen2013pynbody} to compute the radial profiles of the physical parameters. We derive the radial profiles of the input parameter profiles for $R >$ 1 kpc in 25 logarithmic bins. We show the face-on projections of stellar and gas distribution of our sample of barred and unbarred galaxies from \texttt{TNG50} in Figures \ref{fig:barred_star} and \ref{fig:unbarred_star}, respectively.

\begin{figure*}
\resizebox{170mm}{230mm}{\includegraphics{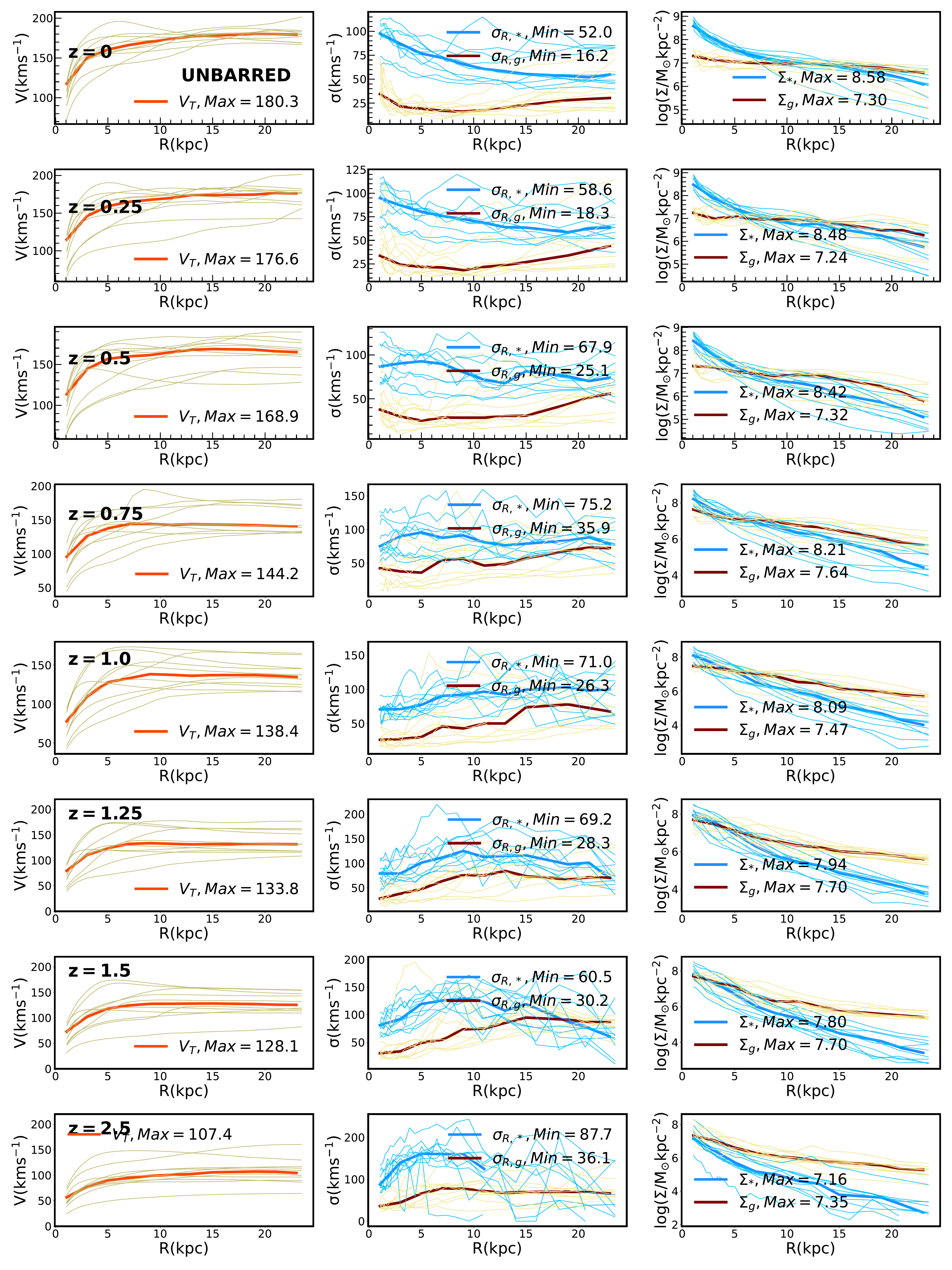}}\\
\caption{Input parameters used to assess the stability of the unbarred subset of Milky Way analogs against local gravitational instabilities at various redshifts. In the first panel, we present the total circular velocity. In the second and third panels, we show the radial velocity dispersion and surface density of stars and gas, respectively. The thick solid line shows the median profile of each parameter, while the thin lines represent individual subhalos.}
\label{Input_unbarred}
\end{figure*}

\begin{figure*}
\resizebox{170mm}{230mm}{\includegraphics{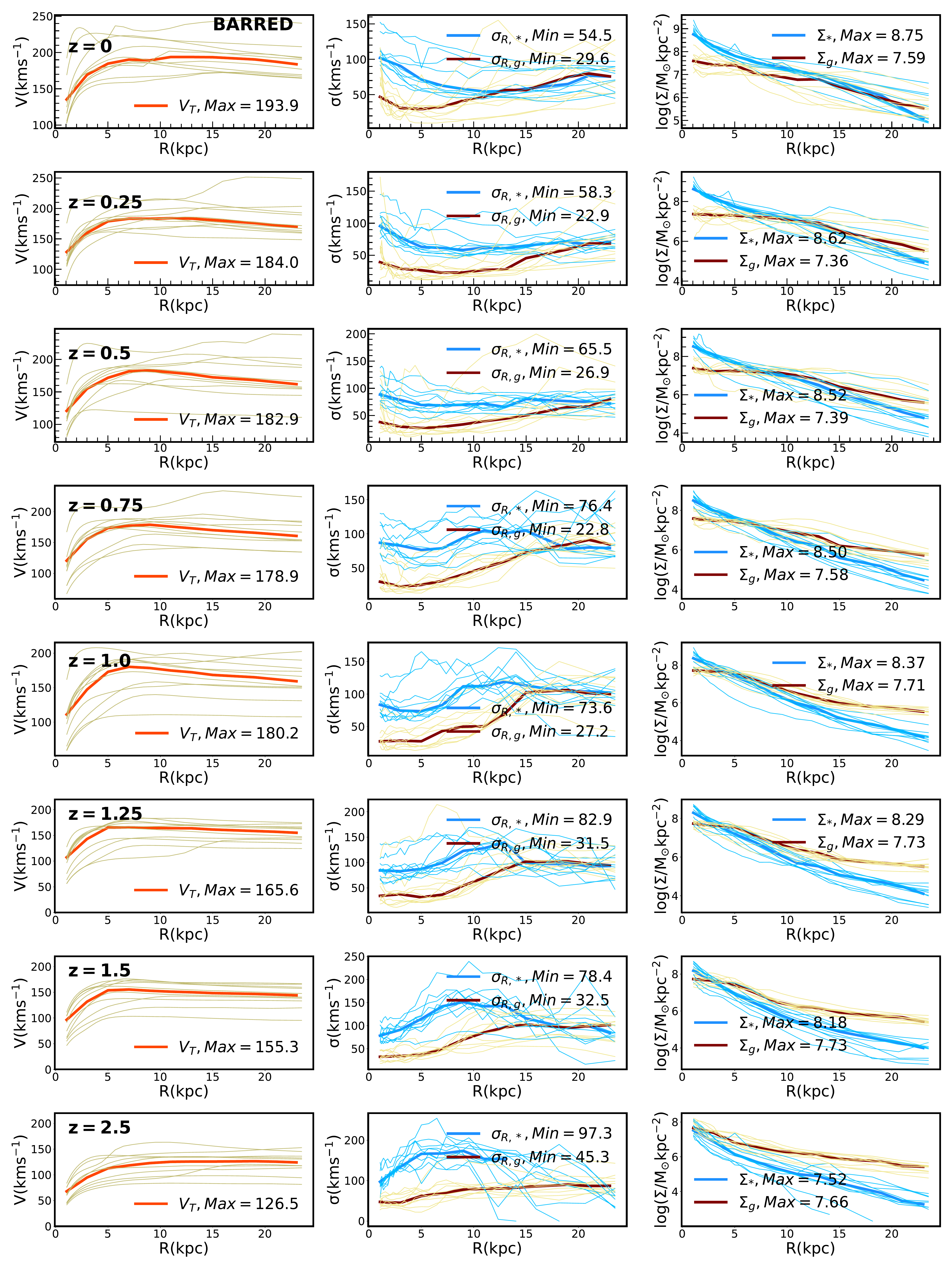}}\\
\caption{Input parameters used to assess the stability of the barred subset of Milky Way analogs against local gravitational instabilities at various redshifts. In the first panel, we show the total circular velocity; in the second and third panels, we present the radial velocity dispersion and the surface density of stars and gas, respectively. The thick solid line shows the median profile of each parameter, while the thin lines represent individual subhalos.}
\label{Input_barred}
\end{figure*}

We show the input parameters for our sample of unbarred and barred galaxies in Figures \ref{Input_unbarred} and \ref{Input_barred}, respectively, at various redshifts. We find that the asymptotic rotation velocity of the MW analogs decreases from $z=0$ to $z=2.5$, indicating that the progenitors of current MWAs did not yet assemble their present-day mass budget at cosmic noon. Both barred and unbarred galaxies have higher radial velocity dispersions at the cosmic noon than today. Also, the radial velocity dispersion of stars is consistently higher than that of gas at any epoch. However, between the barred and the unbarred samples, the barred galaxies have an overall higher velocity dispersion and asymptotic rotation velocity than the unbarred ones. The stellar surface density starts to become greater than or comparable to the gas surface density at $z>1$ in the outer radius. Overall, we find that the progenitors of the current MWAs at cosmic noon have a smaller total mass, typically a higher gas surface density than stars, and a higher radial velocity dispersion than they do at the present epoch.

\section{Dynamical model of local disc instabilities}
We define the dynamical  state of a galaxy using four quantities; 
\begin{enumerate}
\item the net local stability of the galaxy $(Q_{T})$, 
\item critical surface density $\Sigma_{c}$, 
\item time scale for the growth of local instabilities $\tau$ 
\item star formation rate (SFR). 
\end{enumerate}
In this section, we will describe how we compute the first 3 quantities $(Q_{T},\, \Sigma_{c},\, \& \      \tau)$  and we will derive the 
radial profiles of star formation rates from the precomputed star formation rates provided in \texttt{TNG50} MWA catalogs. 
We model the galaxy as a coaxial and coplanar disc of stars and gas interacting gravitationally. Each component is specified
by its radial velocity dispersion, surface density, and angular frequency, the combined system is under the influence of the force field of the dark matter halo. The condition for the stability $(Q_{T})$ of a gravitationally coupled two-component system in equilibrium with the dark matter halo is given by \citep{aditya2024does}
\begin{equation}
\frac{2}{1+Q_{T}^{2}}=\frac{(1-f)}{X_{\star-g}(1+ \frac{(1-f)^{2} q_{\star}^{2}}{4X^{2}_{\star-g}} + R )} +
			\frac{f}{X_{\star-g}(1+ \frac{f^{2} q_{g}^{2}}{4X^{2}_{\star-g}} + R  )}.
\end{equation}
In the above equation, the parameter \( R \) quantifies the stabilizing contribution of the dark matter halo on the two-component star+gas disc and is defined as $R = \frac{\kappa^2_{\mathrm{DM}}}{\kappa^2_{\mathrm{disc}}}$, where \( \kappa^2_{\mathrm{DM}} \) is the epicyclic frequency squared derived from the gravitational potential of the dark matter halo, and \( \kappa^2_{\mathrm{disc}} \) corresponds to that of the combined baryonic disc composed of stars and gas. 
The epicyclic frequency $\kappa$ at a radius R is defined as 
\begin{equation}
\kappa^2(R)= \bigg( R\frac{d\Omega^{2} (R)}{dR} + 4\Omega^{2} (R)  \bigg)
\end{equation}
where $\Omega$ is the angular frequency defined as $\Omega^{2} (R)=\frac{1}{R}\frac{d\Phi}{dR}= \frac{V^{2}}{R^{2}}$, $\Phi$ and $V$ are the gravitational potential and the circular velocity of each component respectively. We use the circular velocity of each component to derive its respective epicyclic frequency. Also, $q_{\star}$ and $q_{g}$ are the classical one-component stability criterion for stars and gas \citep{toomre1964gravitational}, defined as $q_{\star}=\kappa_{disc}\sigma_{R,\star}/\pi G\Sigma_{\star}$ and $q_{g}=\kappa_{disc}\sigma_{R,g}/\pi G\Sigma_{g}$ respectively. The gas fraction is defined as $f = \Sigma_{g}/(\Sigma_{\star}+\Sigma_{g})$,  and $X_{\star-g}=\kappa^{2}_{disc}/[2 \pi G (\Sigma_{\star}+\Sigma_{g})k_{min}]$. $X_{\star-g}$ is the dimensionless wavelength at which it is hardest to stabilize the two-component system. The value of $k_{min}$ for the two-component system is given by 
\begin{equation}
\begin{aligned}
k^{3}(4\sigma_{R,\star}^{2} \sigma_{R,g}^{2}) - 3k^{2}(2\pi G\Sigma_{\star} \sigma_{R,g}^{2} +2\pi G\Sigma_{g} \sigma_{R,\star}^{2}) \\
+2k \kappa^{2}_{net}(\sigma_{R,g}^{2}+\sigma_{R,\star}^{2})-(2\pi G\Sigma_{\star}+2\pi G\Sigma_{g}) \kappa^{2}_{net}=0,
\end{aligned}
\end{equation}
where, $\kappa^{2}_{net}=\kappa^{2}_{disc} + \kappa^{2}_{DM}$ and $\kappa^{2}_{disc}=\kappa^{2}_{\star} + \kappa^{2}_{g}$. We use the values of $\Sigma_{\star}$, $\Sigma_{g}$, $\sigma_{R,\star}$ and $\sigma_{R,g}$, in conjunction with circular velocities of stars, gas and dark matter to estimate $\kappa$ of each component and then $k_{min}$. We then finally proceed to compute $X_{\star-g}$ and then $Q_{T}$. For more details, see \cite{jog1996local,aditya2024does}. The $Q_{T}$ parameter used in this paper is conceptually similar to that introduced by \citet{jog1996local}, but differs in two important ways. First, the formulation adopted in \citet{aditya2024does} explicitly incorporates the contribution of the dark matter halo into the equations governing the growth of perturbations. It derives a dispersion relation and stability criterion that link the dark matter halo potential and gas fraction to the net stability of the galaxy.
As a result, $Q_{T}$ separates the contributions of the disc and dark matter halo, allowing us to quantify the effect of the dark matter halo on the overall stability level.
Second, the most unstable wavenumber, $k_{\min}$, is determined semi-analytically. It is also important to note that $Q_{T}$ does not take into account the vertical structure
of the disc, which has a significant stabilizing effect (e.g., \cite{romeo2011effective,romeo2013simple}). However, understanding the effect of finite vertical thickness on
the stability criterion is relatively straightforward. For a disc of total thickness equal to $2h$ and the thickness is small compared to the wavelength of perturbations,
the finite thickness leads to a reduction of the radial force in the midplane by $(1-e^{-kh})/kh$ \citep{toomre1964gravitational,jog1984galactic,elmegreen2011gravitational}.
The effect could be thought of as a reduction in the overall surface density of the stellar and gas component $\Sigma_{\star}\rightarrow \Sigma_{\star}(1-e^{-kh_{\star}})/kh_{\star}$
and $\Sigma_{g}\rightarrow \Sigma_{g}(1-e^{-kh_{g}})/kh_{g}$.

\cite{lin1993protostars,wang1994gravitational, 2003MNRAS.346.1215B, burkert2013dependence} show that the star formation occurs above a gas surface density called the critical gas surface density.
The critical gas density is defined as \citep{wang1994gravitational,2003MNRAS.346.1215B} 
\begin{equation}
\Sigma_{c}=\gamma \frac{\kappa_{net} \sigma_{z,g}}{\pi G} \, \rm and \,   \gamma=\bigg(1+ \frac{\Sigma_{\star}\sigma_{z,g} }{\Sigma_{g}\sigma_{z,\star} }   \bigg)^{-1}.
\end{equation}

The time scale for the growth of gravitational instabilities, which measures how quickly the gas is converted into stars, is given by \citep{1975ApJ...197..551T, leroy2008star, wong2009timescale}
\begin{equation}
    \tau = \frac{2 \pi}{\frac{\pi G \Sigma_{g}}{\sigma_{z,g}} ( 1 + \frac{\Sigma_{\star} \sigma_{z,g} }{\Sigma_{g} \sigma_{z,\star}})}.
\end{equation}

\section{Results}
With all the ingredients and machinery in place for computing the local stability, critical gas density, and time scale for growth of instabilities, we will present how these quantities vary across the MWAs from cosmic noon (z=2.5) to the present day (z=0).
\subsection{How stability varies across the galaxy disc ?}
We present the radial profile of the stability parameter $(Q_{T})$ against the growth of local axisymmetric instabilities in Figure 5.  
We divide our sample into two subsets of barred and unbarred galaxies to assess if the presence of a bar has any impact on local stability levels. We observe that the MWAs are dynamically stable against the growth of local axisymmetric instabilities $(Q_{T}>2)$ from $z=2.5$ to $z=0$. The minimum value of $Q_{T}$ decreases at lower redshifts, approaching $Q^{min}_{T} \approx 2$, especially in the case of the barred galaxies. Moreover, between barred and unbarred MW analogs, barred analogs consistently have a smaller $Q^{min}_{T}$ at all epochs. The net stability profiles, represented by $Q_{T}$, closely follow those of the gas disc $(Q_{g})$ across all redshifts for both barred and unbarred samples, except between z=0 and z=0.5 in barred galaxies, where the stellar disc dominates the overall stability. The progenitors of Milky Way analogs at $z=2.5$ have not yet assembled their full mass budget and remain significantly less massive compared to their present-day counterparts. While the gas surface density is higher at this epoch, the resulting destabilizing effect is offset by substantially higher radial velocity dispersions and additional dynamical support from the dark matter halo, which together act to stabilize the MWAs. The stellar disc remains stable throughout the redshift range $z=0$, as the increase in surface density over time is counteracted by its consistently high velocity dispersion. It is important to note that IllustrisTNG tends to
produce dark matter-dominated systems, with stellar-to-total enclosed mass ratios that are at least a factor of two lower than those inferred for observed galaxies \citep{marasco2020massive}. Furthermore, \cite{romeo2020lenticulars} show that simulations that underpredict the stellar-to-halo mass fraction produce discs that are
overly gravitationally stable.

\begin{figure*}
\hspace{-1.5cm}
 \begin{tabular}{cc}
    \resizebox{190mm}{90mm}{\includegraphics{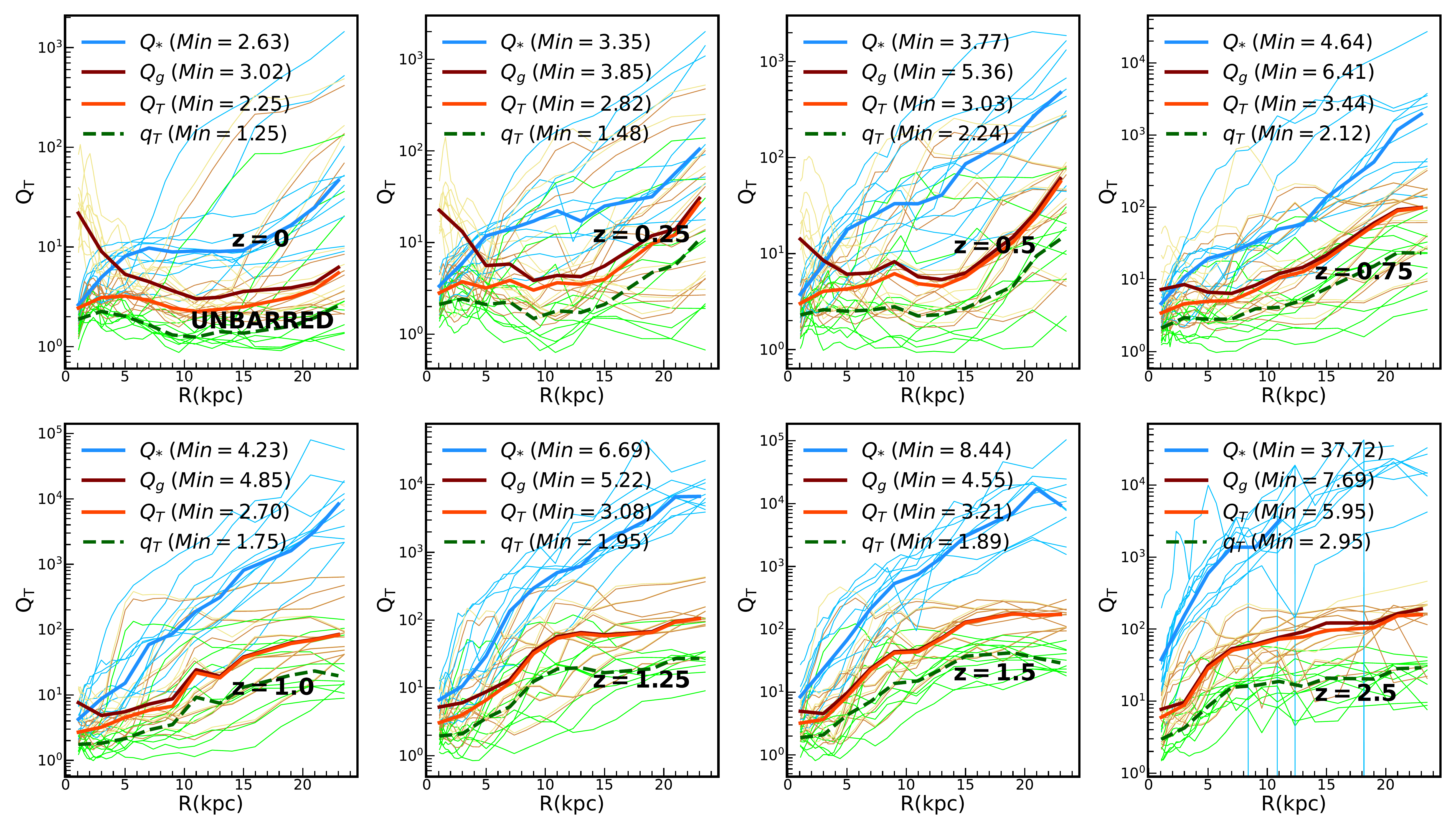}} \\
    \resizebox{190mm}{90mm}{\includegraphics{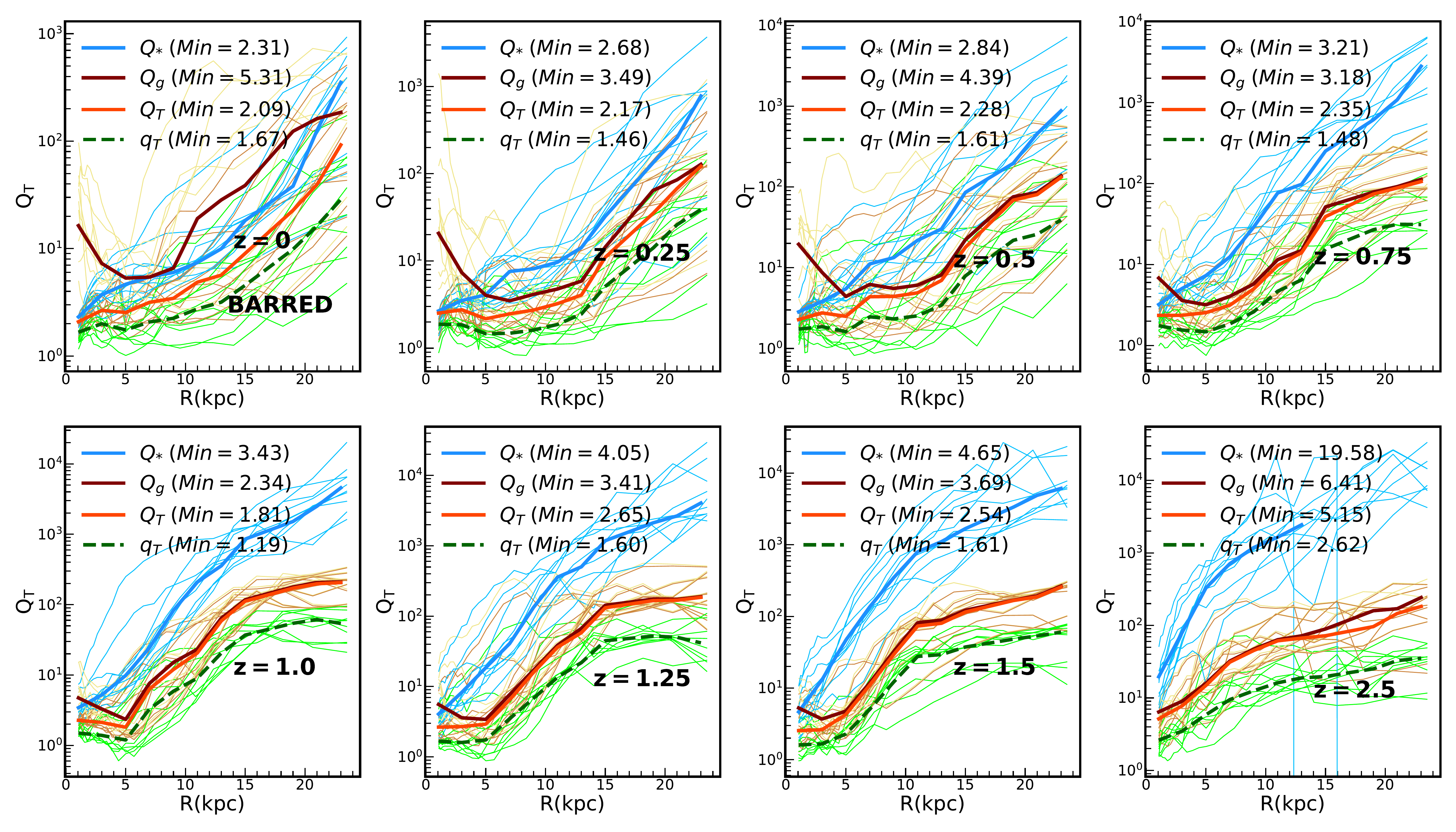}}
    \label{fig:Q_Ubar_Bar}
  \end{tabular}
\caption{We show the radial variation net stability levels of the galaxy $(Q_{T})$, and the stability of just stars $(Q_{\star})$ and gas $(Q_{g})$ across the disc of MWAs. The green curve depicts the stability of galaxies by excluding the stabilizing effect of the DM halo. In the top two rows, we show our results for unbarred MW analogs; in the bottom 
two panels, we show the results for the barred subsets.}  
\end{figure*}

\subsection{Self-regulation of stability levels}
Equation (1) allows us to quantify the effect of the dark matter halo on the net stability of the disc. A value of $R=0$ in equation (1) turns off the stabilizing effect of the dark matter halo, and the net stability of the galaxy is simply due to the combined self-gravity of stars and gas only. From Figure 5, we can see that the  $stellar+ gas$ disc remains stable $Q_{T}>1$ even after excluding the dark matter's contribution at all epochs. This indicates that surface densities and the velocity dispersion of stars and gas self-regulate, such that the galaxy is stabilized against axisymmetric instabilities, even without the contribution of the dark matter halo. These results emphasize that the stellar and gas components can self-regulate the overall stability levels starting from  $z=2.5$. The self-regulation is observed in both the barred and 
unbarred galaxies. Self-regulation of gravitational instabilities has been verified extensively 
for local galaxies spanning diverse morphologies, e.g., see \cite{romeo2013simple,romeo2017drives,romeo2018angular, 
romeo2023specific, aditya2023stability} and for self-regulation of instabilities in hydrodynamical simulations, see 
\cite{renaud2021giant} and \cite{ejdetjarn2022giant}.

\subsection{How does the critical gas surface density vary across the galaxy ?}
The critical gas surface density, \(\Sigma_{c}(R)\), provides an alternative measure of the net gravitational stability of galactic discs. The disc becomes locally unstable whenever the actual gas surface density, \(\Sigma_{g}(R)\), exceeds this critical threshold. In Figure 6, we present the radial profiles of both \(\Sigma_{c}(R)\) and \(\Sigma_{g}(R)\). It is evident that the galaxies remain sub-critical
at all redshifts, i.e. $\Sigma_{g}(R)$ consistently lies below $\Sigma_{c}(R)$. This difference is especially pronounced in the outer regions, where the measured gas surface density drops more steeply than the critical threshold. While both $\Sigma_{c}$ and $\Sigma_{g}$ increase with redshift, the enhanced dynamical support due to an increase in the velocity dispersion ensures that the discs remain stable against local gravitational instabilities. Notably, in barred galaxies, the central values of \(\Sigma_{g}\) approach \(\Sigma_{c}\) more closely than in unbarred systems, consistent with the relatively lower values of 
$Q^{min}_{T}$ in the barred MWAs.

\begin{figure*}
\hspace{-1.5cm}
 \begin{tabular}{cc}
    \resizebox{190mm}{90mm}{\includegraphics{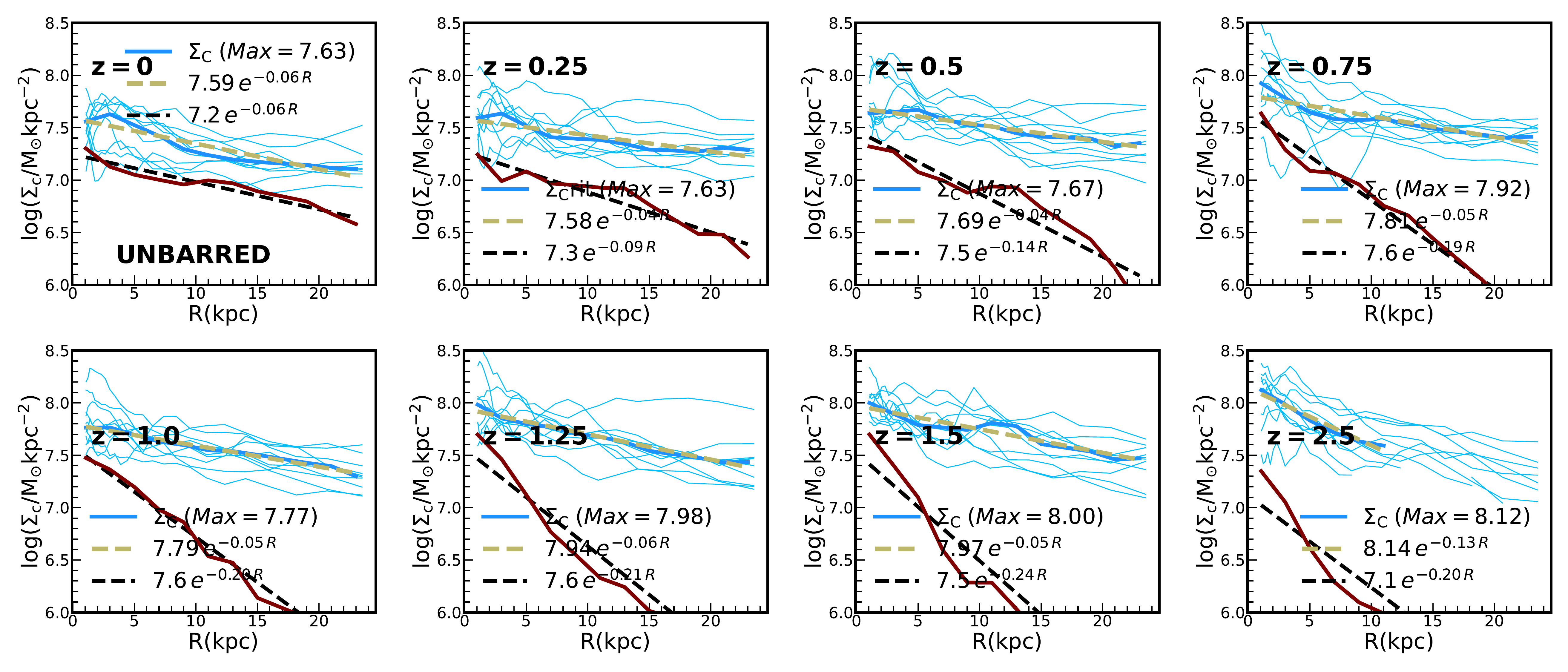}} \\
    \resizebox{190mm}{90mm}{\includegraphics{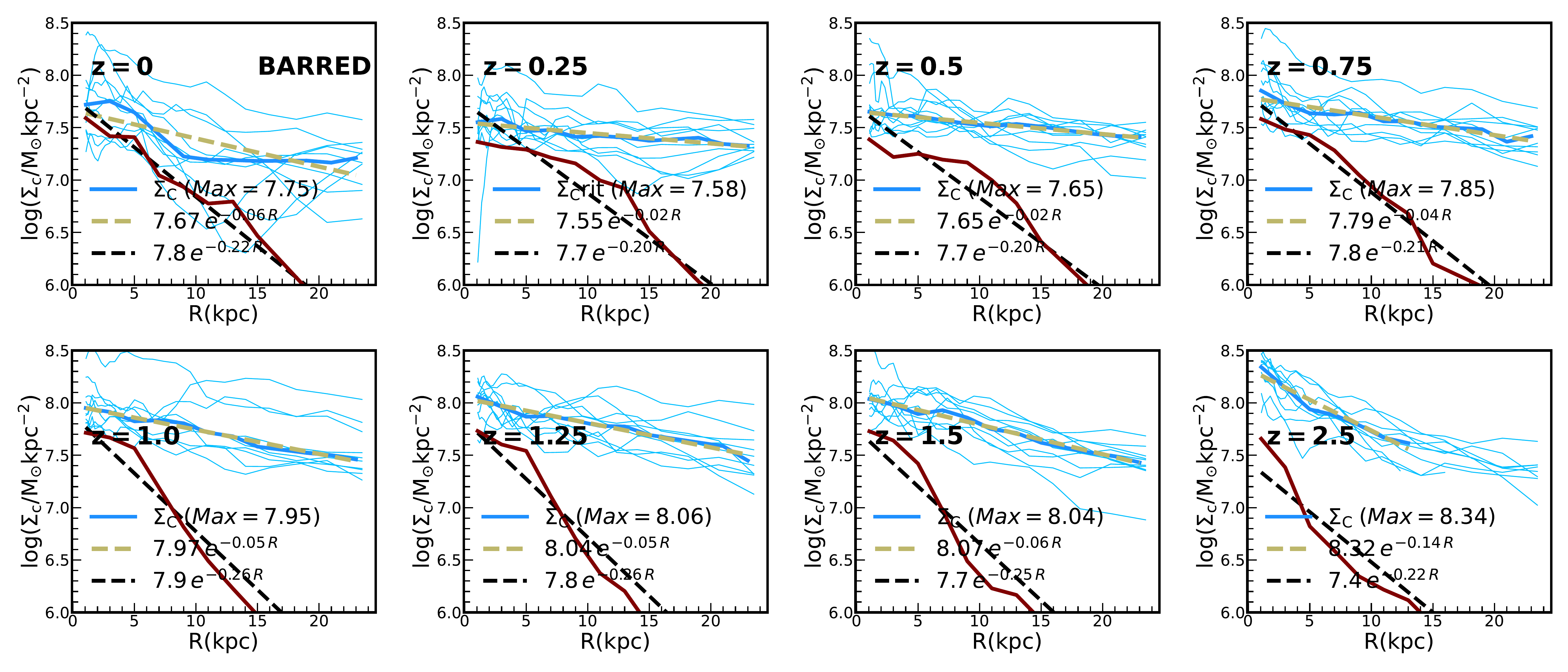}}
  \end{tabular}
\caption{We show the radial variation of critical gas density $(\Sigma_{c})$ of MW analogs (blue), and the measured gas surface density (crimson). The dashed lines indicate exponential fits to 
the critical and the measured gas density. In the top two rows, we show our results for unbarred MW analogs; in the bottom two panels, we show the results for the barred subsets.}  
\end{figure*}

\subsection{How does the timescale for the growth of instabilities vary across galaxy ?}
In Figure 7, we show the radial variation of the timescale for growth of instabilities, $\tau(R)$, for Milky Way analogs across a range of redshifts. 
We find that $\tau$ increases systematically with galactocentric radius, with consistently smaller values in the inner regions and at all epochs, for both barred and unbarred samples. This indicates that local gravitational instabilities can convert gas into stars on timescales of a few Myr in the center. In contrast, star formation via instability-driven collapse can require several Gyr in the outer disc. As redshift increases, the central values of $\tau$ also increase, reflecting the higher velocity dispersions at earlier times, which suppress the rapid growth of instabilities. Between barred and unbarred, barred galaxies exhibit systematically lower values of $\tau$ at all redshifts relative to unbarred systems. Moreover, barred galaxies show a steeper exponential increase in $\tau(R)$, implying that instabilities are strongly confined to the inner few kpc.

\begin{figure*}
\hspace{-1.5cm}
 \begin{tabular}{cc}
    \resizebox{190mm}{90mm}{\includegraphics{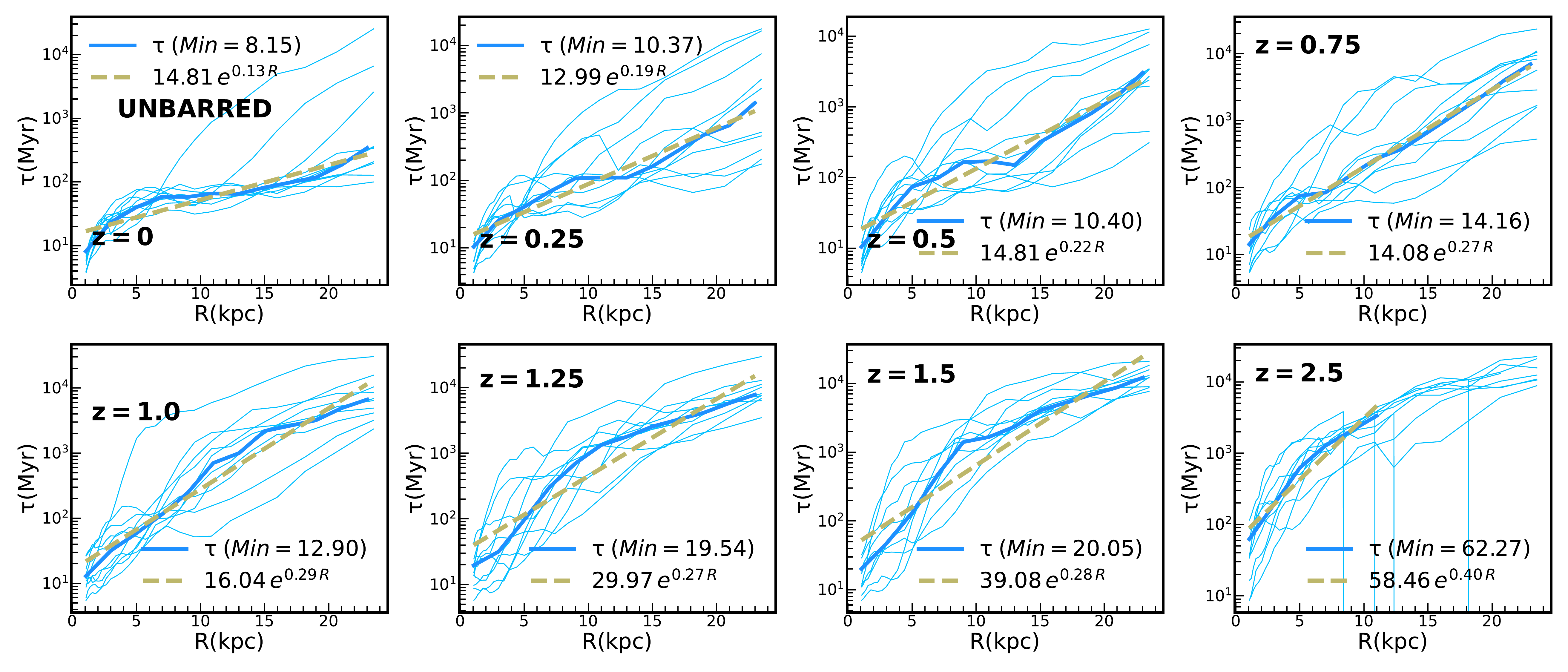}} \\
    \resizebox{190mm}{90mm}{\includegraphics{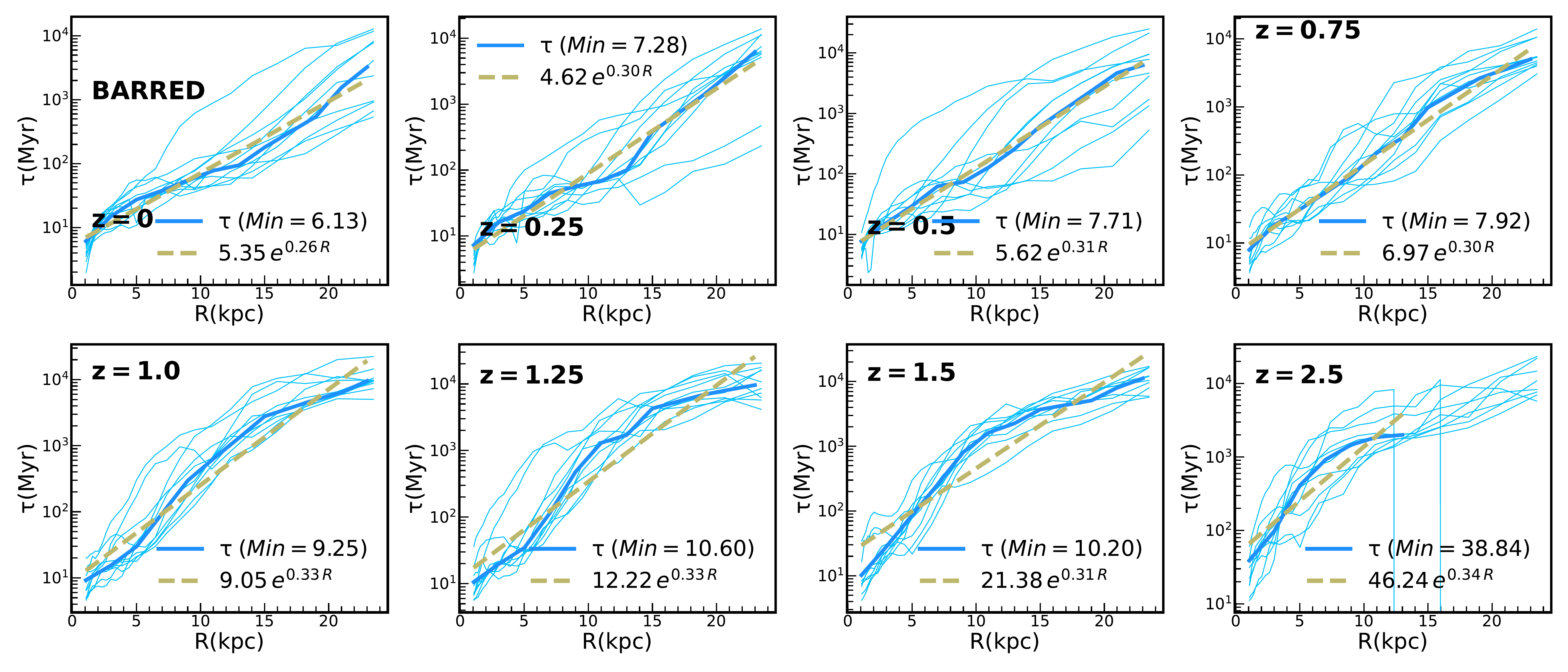}}
  \end{tabular}
\caption{We show the radial variation of time scale for growth of local gravitational instability $\tau$ of MW analogs (blue). The dashed lines indicate exponential fits to $\tau$. In the top two rows, we show our results for unbarred MW analogs, and in the bottom two panels, we show the results for the barred subsets.}  
\end{figure*}

\subsection{How is star formation distributed across the galaxy?}
We show the radial SFR profiles measured directly from the \texttt{TNG50} snapshots in Figure 8. The SFR profile peaks in the central 
region and declines exponentially towards the outskirts. Unbarred galaxies exhibit central SFRs of \(\sim 0.15\)–\(0.28\,M_\odot\,\mathrm{yr}^{-1}\) at \(z \leq 1.25\), which then fall to \(\sim 0.02\)–\(0.05\,M_\odot\,\mathrm{yr}^{-1}\) by \(z = 2.5\). On the other hand barred galaxies maintain higher central SFRs, ranging from \(\sim 0.43\, M_\odot\,\mathrm{yr}^{-1}\) at \(z = 0\) up to \(\sim 0.59\) at \(z = 1.0\), then declining to \(\sim 0.17\) by \(z = 2.5\). The radial distribution of SFR 
remains uniform in the unbarred MWAs at $z=0$ and $z=0.25$, but becomes steeper as we move to higher $z$. However, between barred and unbarred MWAs, barred ones consistently have more centrally concentrated star formation and are characterized by steeper slopes. The centrally concentrated SFR in barred MWAs is consistent with the smaller $Q_{T}$ and steep critical density and instability time scales observed in these galaxies.

\begin{figure*}
\hspace{-1.5cm}
 \begin{tabular}{cc}
    \resizebox{190mm}{90mm}{\includegraphics{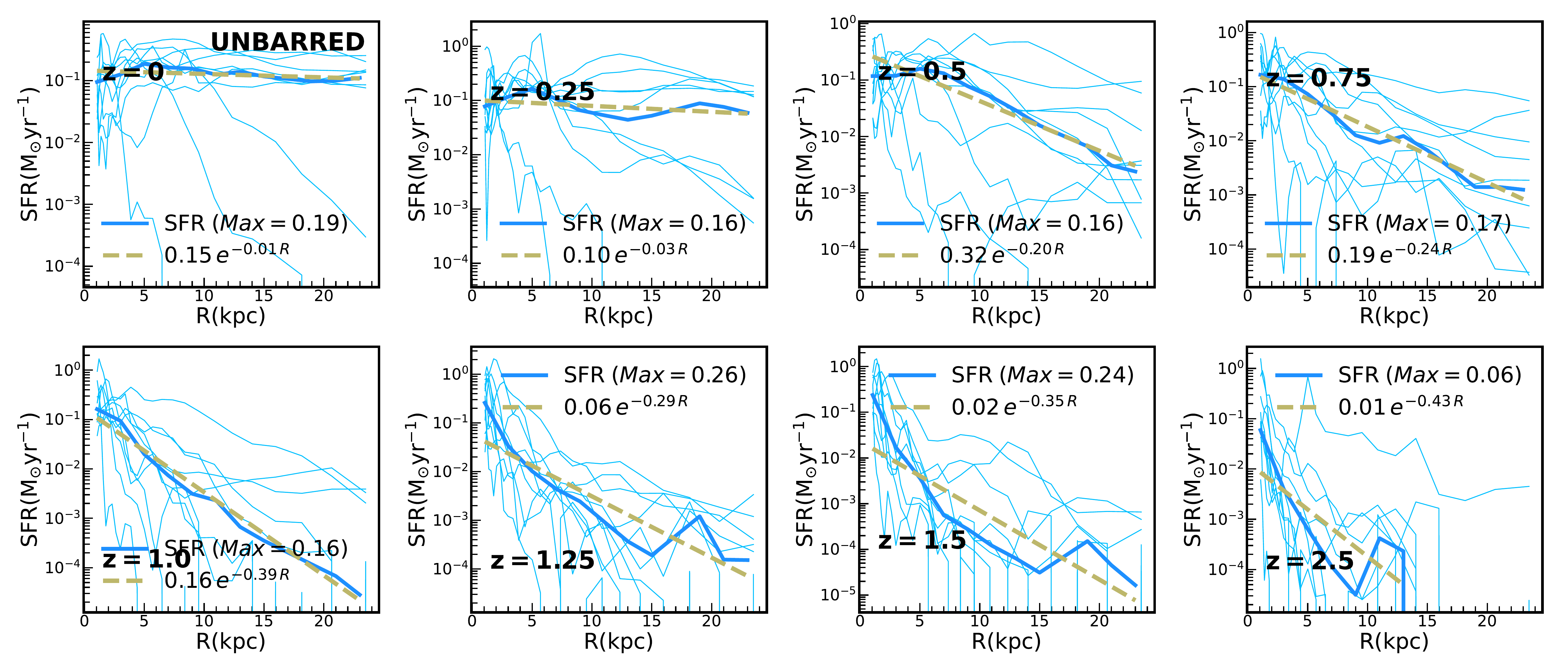}} \\
    \resizebox{190mm}{90mm}{\includegraphics{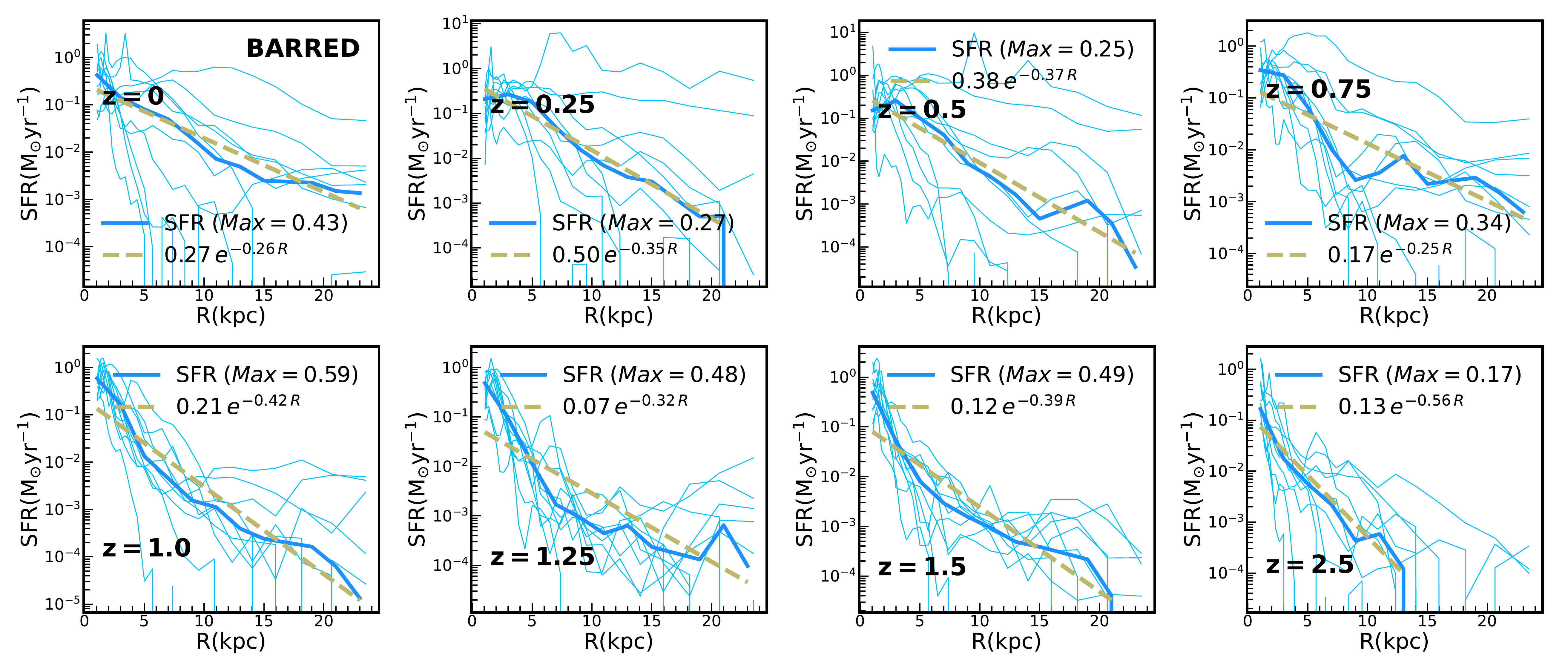}}
  \end{tabular}
\caption{We show the radial variation of star formation rates (SFR) for MWAs (blue). The dashed lines indicate exponential fits to SFR. In the top two rows, we show SFR for unbarred MWAs; in the bottom two panels, we show the SFR for the barred subsets.}  
\end{figure*}

\subsection{Instabilities hiding in plain sight}
In the preceding sections, we observe that the galaxies in \texttt{TNG50} are stable against the growth of local gravitational axisymmetric instabilities at all radii and epochs. However, studies by \cite{romeo2010toomre,2012MNRAS.425.1511H,elmegreen2011gravitational,renaud2018morphology} show that disc galaxies become susceptible to 
instabilities even when $Q_{T}>1$, through gas dissipation and turbulence in ISM at much smaller scales. \cite{elmegreen2011gravitational}, show that the gas dissipation 
modifies the pressure term in the gas disc $\sigma^{2}_{g}k^{2}$ by a factor of $\omega/(\omega + \delta \sigma_{g} k)$, where $\delta^{-1}$ is the time scale for dissipation as a fraction of crossing time. E.g. $\delta=0.5$, mean 2 crossing times. The stability function, including the gas dissipation \citep{elmegreen2011gravitational}, is given by  
\begin{equation}
F_{\delta}(k,\omega) \;=\; 
\frac{2 \pi G \, \Sigma_g \, k}
     {\kappa^2_{net} + \omega^2 + k^2 \sigma_{R,g}^2 \left( \tfrac{\omega}{\omega + \delta k \sigma_R,g} \right)}
\;+\;
\frac{2 \pi G \, \Sigma_{\star} \, k}
     {\kappa^2_{net} + \omega^2 + k^2 \sigma_{R,\star}^2}.
\end{equation}

The above equation assumes that the disc is razor thin and uses the two-fluid approximation \citep{jog1984galactic,jog1996local}. The two-fluid stability function \citep{jog1984galactic,jog1996local,aditya2024does} is obtained by setting $\delta=0$ and then $\omega=0$

\begin{equation}
F_{2f}(k) \;=\;
\frac{2 \pi G \, \Sigma_g \, k}{\kappa^2_{net} + k^2 \sigma_{R,g}^2}
\;+\;
\frac{2 \pi G \, \Sigma_{\star} \, k}{\kappa^2_{net} + k^2 \sigma_{R,s}^2}.
\end{equation}

The effect of gas dissipation can be readily understood by comparing the $F_{\delta}(k)$ and $F_{2f}(k)$, at their marginal stability $(\omega=0)$. The stability function 
$F_{\delta}(k)$ at $\omega=0$ is just

\begin{equation}
F_{\delta}(k) \;=\; 
\frac{2 \pi G \, \Sigma_g \, k}
     {\kappa^2_{net} }
\;+\;
\frac{2 \pi G \, \Sigma_{\star} \, k}
     {\kappa^2_{net} + \omega^2 + k^2 \sigma_{R,\star}^2}.
\end{equation}

At $\omega = 0$, the two-fluid disc loses its pressure support from the gas because of dissipation, and the stabilization of the gas component is only by rotation. In other words, gas dissipation removes 
the pressure contribution in the denominator, so the destabilizing self-gravity terms in the numerator acquires a larger weight. Specifically, since  
\begin{equation*}
\frac{2 \pi G \, \Sigma_g \, k}{\kappa^2_{\mathrm{net}}}
\;>\;
\frac{2 \pi G \, \Sigma_\star \, k}{\kappa^2_{\mathrm{net}} + k^2 \sigma_{R,\star}^2},
\end{equation*}

\begin{equation*}
F_{\delta}(k) > F_{2f}(k).
\end{equation*}
Thus, when gas pressure support is lost through dissipation, the enhanced self-gravity makes the disc unstable. 

In a different scenario, \cite{romeo2010toomre, 2012MNRAS.425.1511H} showed that a galaxy can become unstable if the surface density and 
the velocity dispersion of the gas component are scale dependent even when the galaxy is stable against axisymmetric instabilities.
\cite{romeo2010toomre,2012MNRAS.425.1511H} phenomenologically model the turbulence in the ISM using the Larson type scaling relation given by
\begin{equation*}
\Sigma^{'}_{g}=\Sigma_{g}\left(\frac{k}{k_{0}}\right)^{-a} \&\
\sigma^{'}_{g}=\sigma_{R,g}\left(\frac{k}{k_{0}}\right)^{-b} ,
\end{equation*}
and show that the galaxy is unstable for $b> (a+1)/2 $ and $-2<a<1$, even when $Q_{T}>1$.
The stability function incorporating turbulent ISM can be written as
\begin{equation}
F_{turb}(k) \;=\;
\frac{2 \pi G \, \Sigma_{g} \, k^{1-a} k_0^{a}}
     {\kappa^2_{net} + \sigma_{R,g}^2 \, k^{2-2b} k_0^{2b}}
\;+\;
\frac{2 \pi G \, \Sigma_{\star} \, k}
     {\kappa^2_{net} + k^2 \sigma_{R,\star}^2} .
\end{equation}
As an example, for $a=-1$ and $b=1$, the self-gravity term in the numerator scales as $\sim k^{2}$ (instead of $\sim k$ in the standard $F_{2f}$), indicating that the turbulence in the ISM can drive local disc instabilities even when the $Q_{T}>1$. 
We show the stability functions $F_{2f}(k)$, $F_{\delta}(k)$, and $F_{turb}(k)$ for our sample of MWAs in Figure 9 to illustrate the impact of gas dissipation and turbulence in the ISM. The functions are plotted at marginal stability, i.e., $\omega = 0$. For the scale-dependent
case of $F_{turb}(k)$, we focus on the special choice $a=-1$ and $b=1$, where the contribution from self-gravity scales as $\sim k^{2}$, in contrast to 
the $\sim k^{1}$ dependence in $F_{2f}(k)$ and $F_{\delta}(k)$. We derive the stability functions shown in Figure 9 using the input parameters that 
yield the minimum value of $Q_{T}$. Figure 9 shows that the stability functions have similar behavior at small k or large scale but 
diverge from each other at small scales. Consistent with the discussion in the preceding paragraphs, we can see that $F_{\delta}(k)$ 
is higher than $F_{2f}(k)$, since the energy in the gas component has completely dissipated and the gas component is now supported only through rotation, compared to rotation + dispersion support for the gas component in the two-fluid case. So, the self-gravity 
now has to counteract only the rotation to destabilize the galaxy. So at all epochs, the gas dissipation can effectively destabilize the 
disc at smaller scales, even when the disc is stable against axisymmetric instabilities. In the second scenario, we plot the stability function 
corresponding to the turbulent ISM with $a=-1$ and $b=1$, and use $k_{0}=1$ kpc$^{-1}$. We can clearly see that $F_{turb}$ for $a=-1$ and $b=1$ can destabilize the disc at small scales, more strongly than gas dissipation, and the standard two-fluid case, since the self-gravity of the gas component now scales as $k^{2}$. Furthermore, we observe that $F_{2f}(k)$ for barred galaxies is atleast equal to or greater than the unbarred ones, but we do not see a common trend in $F_{\delta}(k)$ and $F_{turb}(k)$ for barred and unbarred galaxies. Thus, while the MWAs may appear stable against axisymmetric instabilities, as indicated by their large values of $Q^{min}_{T}$, the effects of gas dissipation and turbulence in the ISM can nevertheless continue to drive instabilities at small scales.

\begin{figure*}
\resizebox{190mm}{100mm}{\includegraphics{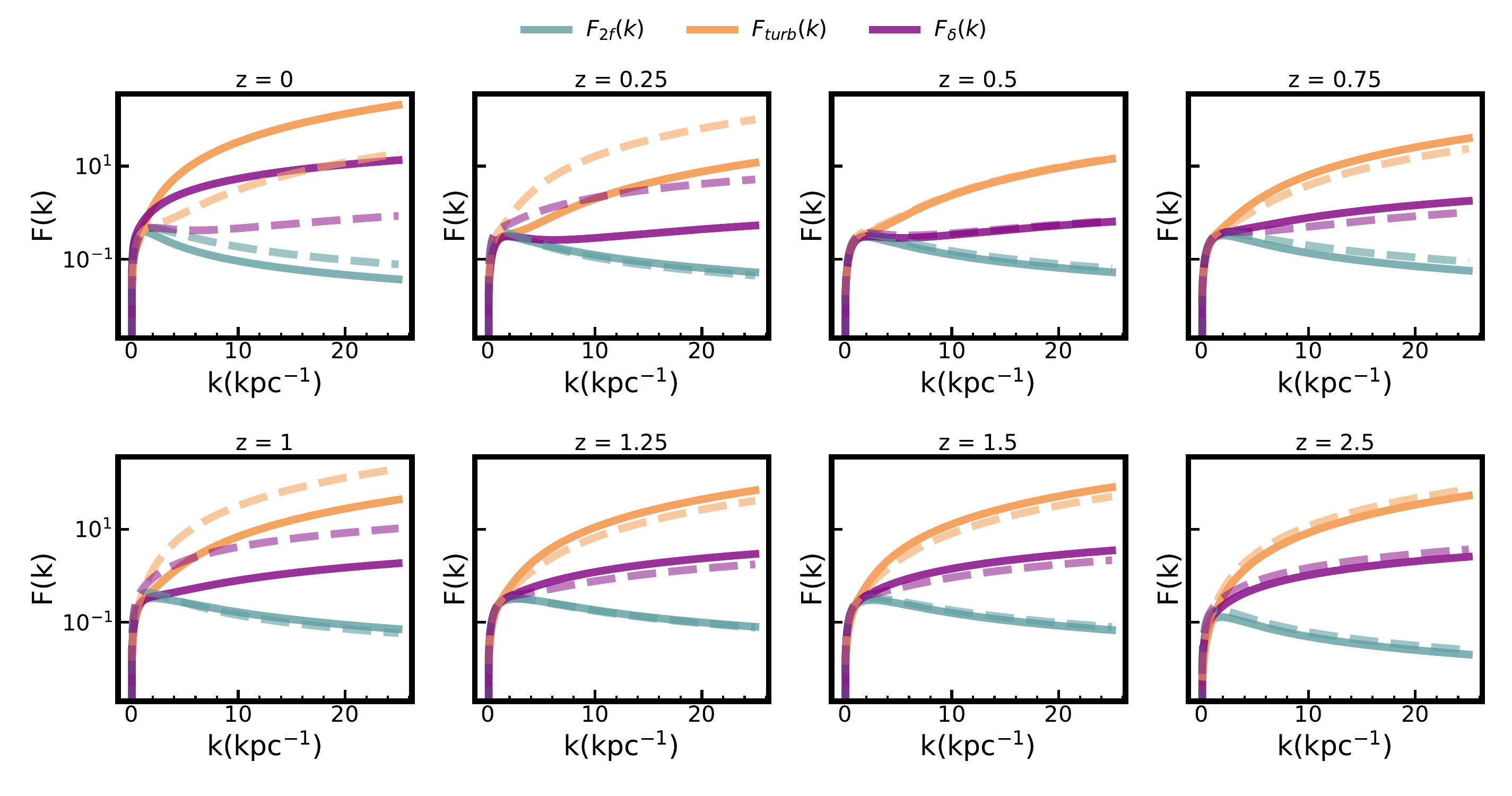}} 
\caption{The stability function $F_{2f}(k)$, $F_{\delta}(k)$ and $F_{turb}(k)$ corresponding to the input parameters that give minimum 
value of $Q_{T}$. The $dashed$ lines correspond to the barred galaxies and $solid$ lines depict the unbarred MWAs.}  
\end{figure*}

\subsection{Tracking $Q_{T}$ from cosmic noon to the present day}
\begin{figure*}
\resizebox{190mm}{70mm}{\includegraphics{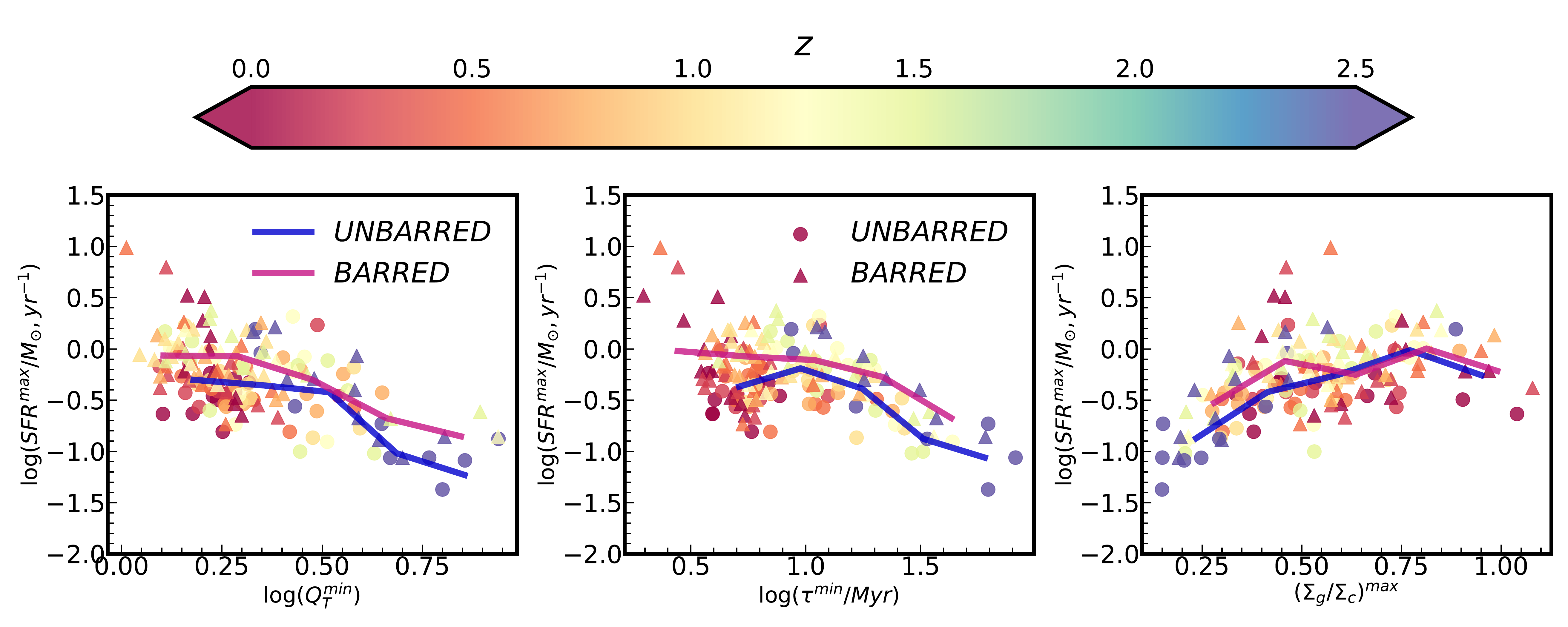}}
\caption{We show the evolution of the star formation rate as a function of $Q_{T}^{\min}$, $\tau^{\min}$, and $(\Sigma_{g}/\Sigma_{c})^{\max}$ from cosmic noon to the present day.
The points are colour-coded by redshift, with triangles and circles representing barred and unbarred galaxies, respectively.
The running median for the barred galaxies is shown by the solid red line, while that for the unbarred galaxies is shown by the solid blue line.}
\end{figure*}

Figure 10 presents a unified evolutionary diagram showing the maximum star formation rate as a function of the minimum stability parameter $Q_T^{\min}$, the minimum instability growth timescale $\tau^{\min}$, and the maximum gas-to-critical surface density ratio $(\Sigma_g/\Sigma_c)^{\max}$, tracing the evolution from cosmic noon to the present day.
The $\mathrm{SFR}_{\max}$ shows an overall tendency to decrease with increasing $Q_T^{\min}$ and $\tau^{\min}$, and to increase with $(\Sigma_g/\Sigma_c)^{\max}$.
While $\mathrm{SFR}_{\max}$ increases as $(\Sigma_g/\Sigma_c)^{\max}$ approaches unity, most galaxies remain subcritical, implying that local axisymmetric instabilities are not the sole trigger of star formation. We also find that, at fixed values of $Q_T^{\min}$, $\tau^{\min}$, and $(\Sigma_g/\Sigma_c)^{\max}$, barred galaxies tend to
exhibit higher star formation rates. At higher redshifts, galaxies tend to occupy regimes of higher $Q_T^{\min}$, $\tau^{\min}$, despite their larger gas fractions. This reflects the increased dynamical support provided by higher gas and stellar velocity dispersions, as well as the stabilizing effect of the dark matter halo. Below $z \sim 1$, the evolution of $\mathrm{SFR}_{\max}$ with respect to $Q_T^{\min}$, $\tau^{\min}$, and $(\Sigma_g/\Sigma_c)^{\max}$ becomes nearly flat, indicating that the star formation tend to self-regulate starting from $z=1$. The connection between star formation and disc instabilities is highly non-trivial, because the star formation is controlled by
two cycles, the \enquote{Alessandro Romeo Cycle (ARC)} \citep{romeo2017drives, Elmegreen2025} driven by disc instabilities shown in Figure 10, and a second cycle driven
by feedback from star formation. The self-regulation of disc instabilities has been extensively studied in
\citep{romeo2020massive, romeo2020lenticulars, romeo2023specific, aditya2023stability}. It is also important to note
here that the quantities $\tau$ and $\Sigma_{c}$ are based on approximate stability derived by \cite{wang1994gravitational} for a
$stars + gas$ system. The \cite{wang1994gravitational} derive their criterion by assuming that the contribution of star and gas to the $F_{2f}(k)$ (equation 7)
are well separated and that each term in $F_{2f}(k)$  peaks at the respective wavenumber of perturbations, instead of the common two-fluid wavenumber. However, despite this shortcoming $\tau$ and $\Sigma_{c}$, based on the approximation by \cite{wang1994gravitational}, provide a qualitative understanding of the ARC and are observationally amenable.

\section{Discussion}
In this section, we will compare the stability levels of MWAs from \texttt{TNG50} with the previous observational studies that have measured stability levels for galaxies in the local and early universe. 

\subsection{$Q_{RW}$ versus $Q_{T}$}
Before comparing the net stability of galaxies in this study with observations from the literature, we first pause to examine the similarities and differences between the various multi-component stability criteria employed in observational works. We use the simple MW model presented in \cite{aditya2024does}. The stellar surface density is described by an exponential profile, \(\Sigma_{\star}(R) = 640\, M_{\odot}\,\mathrm{pc}^{-2} \, e^{-R/3.2\,\mathrm{kpc}}\), where \(\Sigma_{\star,0} = 640\, M_{\odot}\,\mathrm{pc}^{-2}\) is
the central stellar surface density and \(R_D = 3.2\,\mathrm{kpc}\) is the scalelength of the stellar disc \citep{mera1998towards}. The gas surface density 
follows a similar exponential form, \(\Sigma_{g,0}(R) = 28.2\, M_{\odot}\,\mathrm{pc}^{-2} \, e^{-1.65R/(4R_D)}\), where \(\Sigma_{g,0} = 28.2\, M_{\odot}\,\mathrm{pc}^{-2}\) and \(R_{25} = 4R_D\) is the radius at which the B-band surface brightness drops to \(25.5\, \mathrm{mag\, arcsec}^{-2}\) \citep{bigiel2012universal}. The radial velocity dispersion of the exponential disc is modeled as \(\sigma_{R}(R) = \frac{1}{0.6} \sqrt{ \frac{2\pi G R_D \Sigma(R)}{7.3} }\) \citep{leroy2008star, romeo2017drives}.  
The dark matter density is modeled using a pseudo-isothermal halo profile, characterized by a central density \(\rho_{0} = 0.035\, M_{\odot}\,\mathrm{pc}^{-3}\) and a core radius \(R_{c} = 5\, \mathrm{kpc}\) \citep{mera1998towards}. These inputs provide a reference model for evaluating disc stability and enable direct comparison between different stability criteria in 
the literature. \cite{romeo2011effective} derive a 2-component criterion for studying local gravitational instabilities, which was extended to a multi-component case 
by \cite{romeo2013simple}. Both are based on earlier work by  \cite{jog1984galactic,bertin1988global,romeo1992stability,1994A&A...286..799R,jog1996local,rafikov2001local}.
The inverse sum approximation of \cite{wang1994gravitational} is given by
\begin{equation}
    \frac{1}{Q_{WS}} = \frac{1}{Q_{g}}  + \frac{1}{Q_{\star}}.
\end{equation}
The above approximation assumes that the net stability of the galaxy reflects the most unstable component, 
$Q_{WS}= Q_{\star}, \, Q_{g}\xrightarrow{} \infty$ and $Q_{WS}= Q_{g}, \, Q_{\star}\xrightarrow{} \infty$. Thus, by design 
$Q_{WS}$ is smaller than the stability of each component and is also smaller than $Q_{T}$ presented in this work, as shown in Figure 10. The $Q_{RW}$ presented by \cite{romeo2011effective} improves the approximate criterion 
presented by \cite{wang1994gravitational}, and is given by 
\begin{equation}
\frac{1}{Q_{\rm RW}} = 
\begin{cases}
    \dfrac{W_{\sigma}}{T_{\star} Q_{\star}} + \dfrac{1}{T_{g} Q_{g}} & \text{if } T_{\star} Q_{\star} > T_{g} Q_{g} \\
    \dfrac{1}{T_{\star} Q_{\star}} + \dfrac{W_{\sigma}}{T_{g} Q_{g}} & \text{if } T_{\star} Q_{\star} < T_{g} Q_{g}
\end{cases}
\end{equation}
\noindent where the weight function $W_{\sigma}$ is given by
\begin{equation}
W_{\sigma} = \frac{2 \sigma_{\star} \sigma_{g}}{\sigma_{\star}^{2} + \sigma_{g}^{2}},
\end{equation}
\noindent and the thickness correction factors are defined as
\begin{equation}
T_{\star} \approx 0.8 + 0.7 \frac{\sigma_{\star,z}}{\sigma_{\star,R}}, \qquad
T_{g} \approx 0.8 + 0.7 \frac{\sigma_{g,z}}{\sigma_{g,R}}.
\end{equation}
The criterion for stability of galactic disc against axisymmetric instabilities presented by \cite{romeo2011effective}, combines 
the $Q_{\star}$ and $Q_{g}$ through weighted harmonic means and includes the thickness correction by taking into account the 
ratios of the velocity dispersion in the radial and the vertical direction. The $Q_{RW}$ curve is higher than the 
$Q_{T}$ used in this work and as such would provide conservative upper limits on the stability profiles of galaxies (see Figure 10). 
$Q_{T}$ lies in between the inverse sum approximation derived by \cite{wang1994gravitational}
and thickness-corrected weighted harmonic means approximation derived by \cite{romeo2011effective}. The relative difference
between the thickness corrected $Q_{RW}$ and $Q_{T}$, lies between $5\%$ to $10\%$, which was as also shown in earlier studies by
\cite{romeo2013simple}. On the otherhand, the $Q_{WS}$ underestimates $Q_{T}$ by $10\%$ to $15\%$.

\begin{figure}
\resizebox{90mm}{70mm}{\includegraphics{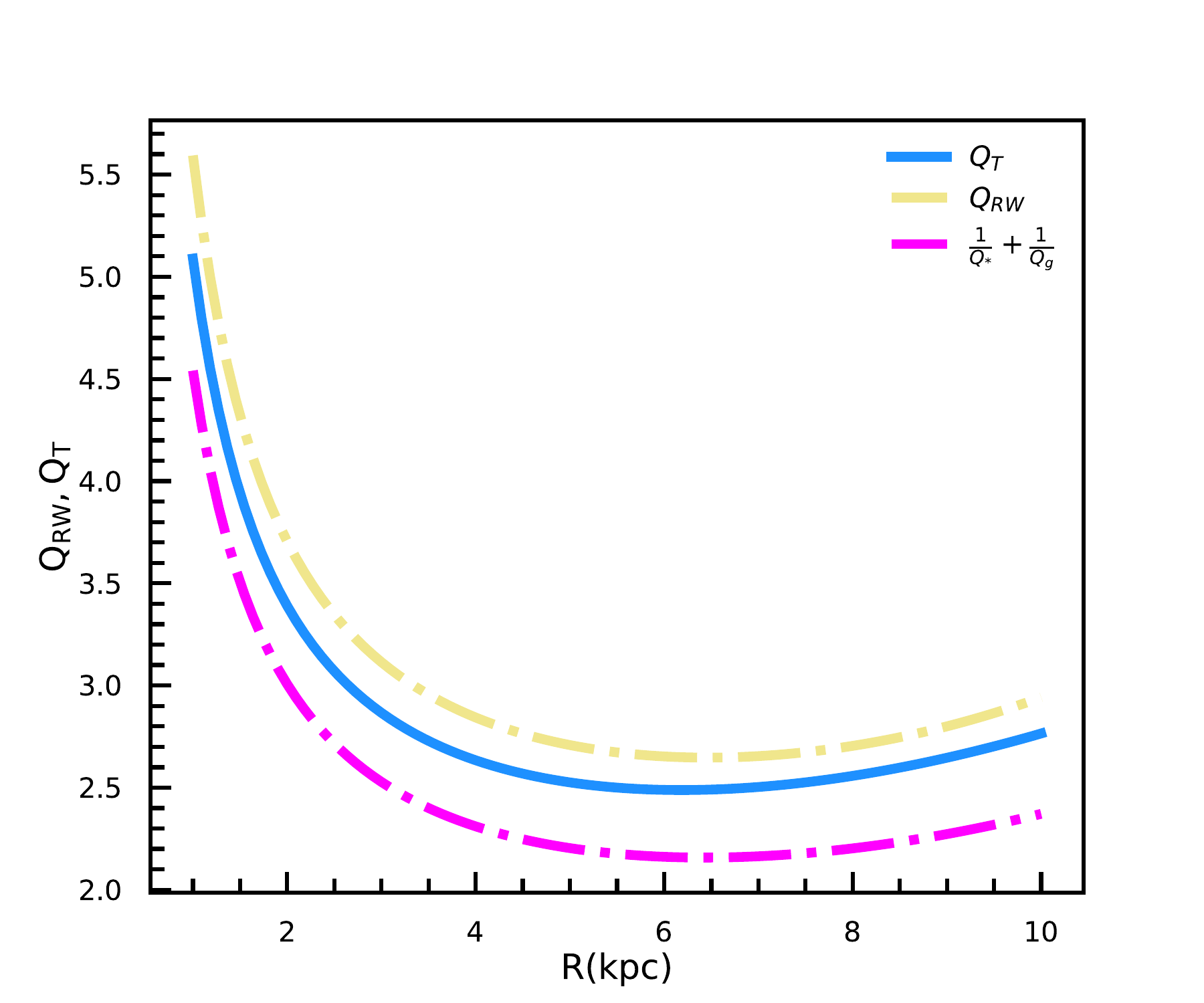}} 
\caption{We compare the stability criterion $(Q_{T})$ used in this work with $Q_{WS}$ and $Q_{RW}$ used in literature.}  
\end{figure}

\subsection{Galaxies at z = 0}
\cite{romeo2011effective,romeo2013simple} studied the stability of 12 local star-forming galaxies observed as part of the THINGS survey \citep{leroy2008star}, and found that nearby spiral galaxies typically have \( Q_{\mathrm{RW}} > 1 \), with a global median value of 2.2, indicating that these galaxies are stable against the growth of axisymmetric perturbations. More recent work by \cite{bacchini20243d} investigated the stability of galaxies over a wide redshift range (\( 0 < z < 5 \)) using a 3D criterion developed by \cite{nipoti2024local}. \cite{bacchini20243d} found no unstable gas discs in the redshift range \( 0 < z < 1 \). It is important to note that the analysis by \cite{bacchini20243d} is a one-component treatment of the gas disc. \cite{aditya2023stability} examined the stability of 175 nearby galaxies from the SPARC catalog \citep{2016AJ....152..157L}, which includes galaxies with diverse physical and morphological properties, and found that 91\% of SPARC galaxies have a \( Q_{RW}^{min} > 1 \). \cite{aditya2023stability} found that only a total of 15 galaxies in the SPARC catalog show \( Q_{RW}^{min} < 1 \), of which 5 are irregulars, and 10 are spirals.  We find that, similar to the observations, MWAs also have $Q^{min}_{T} > 2$, with barred galaxies having a smaller $Q^{min}_{T}$ compared to unbarred ones. Furthermore, \cite{leroy2008star,leroy2013molecular} measures 
the gas depletion time scale equal to $2.2Gyr$ for galaxies in the THINGS sample, which is comparable to our measurement of $\tau$ beyond the central $2kpc$. Similarly, our measurement of $\Sigma_{g}/\Sigma_{c}\approx 0.4$ is comparable to  $\Sigma_{g}/\Sigma_{c}\approx 0.5 - 0.7$ predicted by 
\cite{boissier2003star,aditya2023stability}.

\subsection{Galaxies at cosmic noon (z = 2.5)}
\cite{bacchini20243d} study the stability of gas disc in two galaxies; zC-400569 at $z \approx 2.24$ and zC-488879
$z \approx 1.47$ \citep{2014ApJ...785...75G,2023A&A...672A.106L}. \cite{bacchini20243d} find that the gas disc of zC-400569 is locally 
unstable, however zC-488879 remains stable at all radius. However, we find that none of the MWAs in our study are prone to 
instabilities at $z=2.5$. zC-400569 already has a stellar mass equal to $10^{11}M_{\odot}$ at $z=2.5$, comparable to MW's present day stellar mass. However, none of the MWAs in our sample have assembled a stellar mass comparable to $10^{11}M_{\odot}$ at $z=2.5$. Furthermore, the both zC-488879 and zC-488879 have an observed star formation rate equal to $10^{2}M_{\odot}yr^{-1}$ compared to $0.01 -0.1 M_{\odot}yr^{-1}$ for \texttt{TNG50} MWAs at z=2.5.

\subsection{Beyond cosmic noon}
\cite{bacchini20243d} and \citep{aditya2023h} quantify the stability levels in a sample of 6 dusty star-forming galaxies (DSFGs) observed by 
\citep{rizzo2020dynamically,rizzo2021dynamical}. To reiterate the difference between the analysis, \cite{bacchini20243d} measured the stability of the gas disc using the simple Toomre stability criterion and the 3D criterion presented by \cite{nipoti2024local}, and \cite{aditya2023stability} used the two-component criterion \citep{romeo2011effective} to assess the net stability of the galaxy disc consisting of stars and gas. The DSFGs observed by \cite{rizzo2020dynamically,rizzo2021dynamical} have stellar mass $>10^{10}M_{\odot}$ and star formation rates 
$>10^{2}M_{\odot}yr^{-1}$. \cite{aditya2023stability} finds that the gas component is Toomre unstable in only 2 galaxies out of 6, compared to 5 unstable gas discs in  \cite{bacchini20243d}. However, \cite{aditya2023stability} shows that all six galaxies are susceptible to the growth of local axisymmetric instabilities (\( Q_{\mathrm{RW}} < 1 \)) when the self-gravity of both stars and gas is included, highlighting the importance of a multi-component stability analysis. 

\subsection{ Local Instability ($Q_T$), global Instability (Bars) \& star formation}
From Figure 5, we find that $Q_T$ ($r< 5 \hspace{1mm}$ kpc) of the barred galaxies is close to 2 since z~1.5, given most of the bars in the \texttt{TNG50} form around this epoch \citep{2022MNRAS.512.5339R, Kataria.Vivek.2024}. While for the unbarred galaxies, the $Q_T$ ($r<5 \hspace{1mm}$ kpc) is always larger than that of barred galaxies at all epochs since $z \approx 1.5$ to the present epoch. This agrees with the notion that global non-axisymmetric instabilities become prominent in colder disks having lower velocity dispersion \citep{Ostriker.Peebles.1973,  Kataria.Das.2018,Aditya_2025} and higher surface density\citep{Kataria.Shen.2022,Ansar.et.al.2023,Kataria.Shen.2024,Chen.et.al.2025, Kataria.et.al.2025}. However, the prediction of the star formation rates is underestimated in the region where the local stability criterion ($Q_T$) is close to 1 in the \texttt{TNG50}. This indicates the subgrid resolution for baryonic physics, i.e., star formation and stellar feedback, used in these simulations. The impact of subgrid resolution has been quantified recently for AURIGA galaxies \citep{pakmor2025auriga} and  \texttt{TNG50} galaxies \citep{Yetli.et.al.2025}, which can severely affect the star formation rates by a factor of more than 2.

\subsubsection{What does $Q_{T}\geq 1$ means ?}
The MWA in our sample remains remarkably stable at all radii and all epochs $(z<2.5)$. In the two-fluid stability paradigm, $Q_{T}\geq1$ essentially means 
that the axisymmetric perturbations will continue oscillating. On the other hand, a value of $Q_{T}<1$ is associated with growing/decaying axisymmetric 
perturbations. \cite{inoue2016non}, study formation of clumps in zoom-in cosmological simulations and show that the $Q_{RW}<1$ is confined only to collapsed 
clumps due to high surface density; however, the inter-clump region continues to remain stable $Q>1$. Interestingly, \cite{inoue2016non} find that the proto-clumps
continue to be stable $Q_{RW}=1-3.3$, with some of their models exceeding even $Q_{RW}>3.3$. \cite{inoue2016non}, conclude that the clump formation is inconsistent 
with the standard linear stability analysis, also see \cite{renaud2021giant}. However, \cite{elmegreen2011gravitational}, showed that gas dissipation 
can increase the instability threshold from $Q_{T}=1$ to $Q_{T}=2-3$. So, a value of $Q_{T}=2-3$, is consistent with dissipative instabilities if 
the turbulence decays on timescale similar to the dynamical timescales. \cite{romeo2010toomre, renaud2021giant}, argue that the growth of 
instabilities is scale-dependent and that instabilities can grow on smaller scales, even when $Q_{T}>1$ on larger scales. \cite{romeo2010toomre, hoffmann2012effect}, 
show that when the gas density and dispersion are scale dependent $\Sigma_{g}\propto l^{a}$ and $\sigma_{g}\propto l^{b}$, then the disc remains unstable for $b> (a+1)/2 $ and $-2<a<1$, even when $Q_{T}>1$. Studies by \cite{michikoshi2014pitch,ceverino2015early,inoue2016non} point out that the discs can become unstable through non-linear and violent disc instabilities even when $Q_{T}>1$. Thus, although our sample of MWAs from \texttt{TNG50} are stable against the growth of axisymmetric instabilities, these galaxies may still be susceptible to perturbations driven by gas dissipation, scale-dependent instabilities, and violent non-linear disc processes, which can operate even when $Q_{T} > 1$.

\section{Summary}
In this work, we have quantified the net stability levels of Milky Way analogs from the \texttt{TNG50} against the growth of local axisymmetric 
instabilities. We choose a representative sample of 20 MWAs (10 barred + 10 unbarred)from the \texttt{TNG50} catalog and follow the evolution 
of their dynamical properties from $z=2.5$ to $z=0$. We find that:

\begin{enumerate}
\item The MW analogs in \texttt{TNG50} are stable against local axisymmetric instabilities from cosmic noon $z=2.5$ to the present day. The minimum value of the median $Q^{min}_{T}>2$, with relatively higher values of $Q^{min}_{T}$ at higher redshift. Higher gas velocity dispersion counteracts the increase in gas density at higher redshifts, which can potentially destabilize the galaxy.

\item The barred galaxies in our sample consistently have a smaller $Q^{min}_{T}$ at all epochs. This is consistent with the smaller 
time scale in which instabilities can convert gas into stars and higher star formation rates observed in the barred galaxies. The timescale for the growth of instabilities increases steeply with radius in barred galaxies compared to unbarred ones.

\item The gas density is subcritical at all epochs. So, the local axisymmetric instabilities are not the main channel of star formation. The 
star formation persists despite subcritical densities, with the barred galaxies exhibiting a higher SFR $(0.1 -0.6 M_{\odot}yr^{-1})$,
compared to the unbarred galaxies $(<0.3M_{\odot} yr^{-1})$.

\item The timescale for the growth of instabilities in MWAs rises exponentially. The inner regions can convert gas in a few Myrs compared to Gyrs in the outer disc. This naturally explains the centrally peaked star formation profiles.

\item Although the progenitors of the MWAs at z=2.5 have a higher gas surface density, they also exhibit higher gas velocity 
dispersions and a smaller dynamical mass. Thus, the combined support of gas dispersion and dark matter halo ensures that 
$Q_{T}>>1$, even during the early stages of formation. Thus, the higher gas fraction alone does not guarantee the growth of 
axisymmetric instabilities.

\item Despite removing the contribution of the dark matter halo, the stellar+gas disc remains stable. This implies that the 
stellar+gas disc self-regulates the surface densities and velocity dispersions. The self-regulation mechanism operates in both barred and unbarred galaxies and at all epochs.

\item We study the effect of gas dissipation and turbulence in ISM and find that the gas dissipation and turbulent ISM can destabilize the MWAs even when the galaxies are stable against axisymmetric instabilities.

\item We conclude that the MWAs are stable against growth of axisymmetric instabilities at all epochs $(Q_{T}>1)$; \textit{from cosmic noon to the present day}. However, we note that even when $Q_{T}>1$, these galaxies continue to be unstable through gas dissipation and scale-dependent instabilities driven by turbulence in the ISM. All these processes aid in increasing the threshold instability levels from $Q_{T}<1$ to $Q_{T}=2-3$. 
\end{enumerate}

\section{Acknowledgements}
We thank the anonymous referee for their insightful comments and constructive feedback, which significantly improved the paper. It is a pleasure to thank Alessandro Romeo for insightful and stimulating discussions. The IllustrisTNG simulations were undertaken with compute time awarded by the Gauss Centre for Supercomputing (GCS) under GCS Large-Scale Projects GCS-ILLU and GCS-DWAR on the GCS share of the supercomputer Hazel Hen at the High Performance Computing Center Stuttgart (HLRS), as well as on the machines of the Max Planck Computing and Data Facility (MPCDF) in Garching, Germany. SKK acknowledges the support from the INSPIRE Faculty award (DST/INSPIRE/04/2023/000401) from the Department of Science and Technology, Government of India, and the PARAM Supercomputing facility at IIT Kanpur. 

\newpage
\sloppy
\bibliography{2366_d2.bib}{}
\bibliographystyle{aasjournalv7}

\end{document}